\documentclass[lettersize,journal]{IEEEtran}
\usepackage{etex}

\usepackage{amsfonts}
\IEEEoverridecommandlockouts
\ifCLASSINFOpdf
\else
\fi
\usepackage{booktabs}
\usepackage{makecell}
\usepackage{graphicx}
\usepackage{subfigure}
\usepackage{cite}
\usepackage{color}
\usepackage[pagebackref=false,breaklinks=false,letterpaper=true,colorlinks,citecolor=blue,linkcolor=blue, anchorcolor=blue, bookmarks=false]{hyperref}
\usepackage{amsmath}
\usepackage{amssymb}
\usepackage{array}
\usepackage{bm}
\usepackage{algorithm}

\usepackage{algorithmic}
\usepackage{pifont}
\newcommand{\et}{\emph{et al.}}
\usepackage{orcidlink}
\usepackage{cite}
\usepackage{lettrine} 

\title{A Federated Learning-based Lightweight Network with Zero Trust for UAV Authentication}

\author{Hao Zhang$^{\orcidlink{0000-0003-1923-589X}}$, \IEEEmembership{Member, IEEE}, 
Fuhui Zhou$^{\orcidlink{0000-0001-6880-6244}}$, \IEEEmembership{Senior Member, IEEE}, \\
Wei Wang$^{\orcidlink{0000-0002-2772-4856}}$, \IEEEmembership{Senior Member, IEEE},
Qihui Wu, \IEEEmembership{Fellow, IEEE}, 
and Chau Yuen$^{\orcidlink{0000-0002-9307-2120}}$, \IEEEmembership{Fellow, IEEE}
\thanks{
This work was supported in part by the National Natural Science Foundation of China under Grant 62222107 and the Yangtze River Delta Science and Technology Innovation Community Joint Research (Basic Research) Project under Grant BK20244006. 
The work of C. Yuen was supported by National Research Foundation, Singapore and Infocomm Media Development Authority under its Future Communications Research \& Development Programme FCP-NTU-RG-2024-025

Hao Zhang and Fuhui Zhou are with the College of Artificial Intelligence, Nanjing University of Aeronautics and Astronautics, Nanjing 211106 China. They are also with the Key Laboratory of Dynamic Cognitive System of Electromagnetic Spectrum Space (Nanjing University of Aeronautics and Astronautics) and with the Ministry of Industry and Information Technology, Nanjing, 211106, China (email: haozhangcn@nuaa.edu.cn, zhoufuhui@ieee.org).

Qihui Wu is with the College of Electronic and Information Engineering, Nanjing University of Aeronautics and Astronautics, Nanjing 211106 China. He is also with the Key Laboratory of Dynamic Cognitive System of Electromagnetic Spectrum Space (Nanjing University of Aeronautics and Astronautics) and with the Ministry of Industry and Information Technology, Nanjing, 211106, China (email: wuquhui2014@sina.com).

Wei Wang is with the School of Computer Science, Wuhan University, Wuhan 430072, China (e-mail: wangw@whu.edu.cn).

Chau Yuen is with the School of Electrical and Electronic Engineering, Nanyang Technological University, Singapore 639798 (email: chau.yuen@ntu.edu.sg)
}
}

\begin{document}

\maketitle

\begin{abstract}
Unmanned aerial vehicles (UAVs) are increasingly being integrated into next-generation networks to enhance communication coverage and network capacity. However, the dynamic and mobile nature of UAVs poses significant security challenges, including jamming, eavesdropping, and cyber-attacks. To address these security challenges, this paper proposes a federated learning-based lightweight network with zero trust for enhancing the security of UAV networks. A novel lightweight spectrogram network is proposed for UAV authentication and rejection, which can effectively authenticate and reject UAVs based on spectrograms. Experiments highlight LSNet's superior performance in identifying both known and unknown UAV classes, demonstrating significant improvements over existing benchmarks in terms of accuracy, model compactness, and storage requirements. Notably, LSNet achieves an accuracy of over $80\%$ for known UAV types and an Area Under the Receiver Operating Characteristic (AUROC) of $0.7$ for unknown types when trained with all five clients. Further analyses explore the impact of varying the number of clients and the presence of unknown UAVs, reinforcing the practical applicability and effectiveness of our proposed framework in real-world FL scenarios.
\end{abstract}

\begin{IEEEkeywords}
Unmanned aerial vehicles (UAVs), zero trust, federated learning (FL), security, authentication, rejection.
\end{IEEEkeywords}

\section{Introduction}
\lettrine[lines=2]{R}{ECENT} developments in wireless communication technology have significantly expanded the use and capabilities of wireless devices, leading to an increased demand for such services. Unmanned aerial vehicles (UAVs) are increasingly becoming crucial components of next-generation networks, enhancing communication coverage and boosting the capacity of terrestrial base stations \cite{al2023zero}. Additionally, UAVs are anticipated to play a pivotal role in the architecture of the sixth-generation (6G) networks, particularly in densely populated areas. These UAVs enable the rapid and efficient scalability of the 6G network to accommodate fluctuating demands. They can also provide essential wireless backhaul for small cells or support the broader 6G network infrastructure \cite{zhang2025revolution}. Moreover, UAVs hold promise in fostering the development of more dynamic, engaging, and interactive virtual environments, crucial for powering metaverse applications and services by leveraging their aerial capabilities and processing power. 

However, the integration of UAVs into 6G networks poses significant security challenges. UAVs are vulnerable to various security threats, including jamming, eavesdropping, and cyber-attacks. These threats can compromise the confidentiality, integrity, and availability of the data transmitted by UAVs, thereby undermining the overall security of the network. To address these security challenges, it is essential to develop robust security mechanisms that can protect UAVs from potential security threats. 

\begin{figure}
\centering
\includegraphics[width=0.5\textwidth]{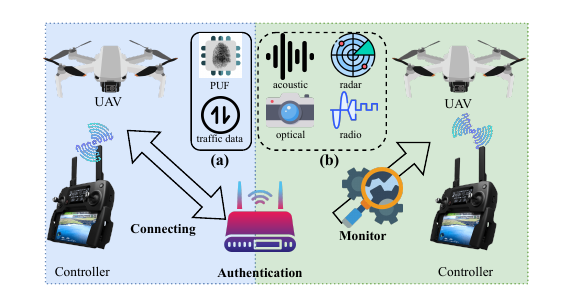}
\caption{The methods of UAV detection and authentication. }
\label{fig:auth}
\end{figure}

As shown in Fig. \ref{fig:auth}, UAV authentication methods can be classified into two categories, namely, traditional authentication mechanisms and non-contact UAV detection techniques. 
Traditional authentication mechanisms, also known as access control mechanisms, identify and verify the identity of UAVs and ground stations by using physical unclonable functions (PUFs) \cite{alladi2020secauthuav,zhang2024prlap}, traffic data analysis \cite{bisio2018unauthorized,alipour2019machine}, and other methods. 
However, traditional access control mechanisms are not well-suited for UAV networks due to their dynamic and mobile nature. Moreover, it is hard to design a lightweight and robust PUF-based authentication mechanism that can adapt to the dynamic and mobile nature of UAV networks. 
 
Apart from these traditional authentication mechanisms, non-contact UAV detection techniques have been proposed to detect and classify UAVs based on their unique features. Physical signals include acoustic \cite{seo2018drone,anwar2019machine}, radar \cite{zhang2017classification}, radio-frequency (RF) \cite{ozturk2020rf,al2019rf,medaiyese2021machine,zhao2018classification,allahham2020deep} signals and camera-based images and videos \cite{thai2019detection,rozantsev2016detecting} that are used to detect and classify UAVs. 
Acoustic-based methodologies often exhibit heightened sensitivity to environmental noise, while the efficacy of visual data from cameras can be compromised by conditions such as obstructive structures, ambient lighting, and other local variables. In contrast, techniques that utilize RF signals and network traffic steam are generally less affected by such environmental changes. However, commercial UAVs typically operate on exclusive communication channels that are protected by various levels of encryption, presenting significant challenges in data acquisition and surveillance. Therefore, it is crucial to optimize the utilization of RF signals for UAV detection in environments susceptible to noise.

Recent works have focused on applying zero-trust security mechanisms to enhance the security of UAV networks. For example, the authors in \cite{haque2024enhancing} proposed a zero-trust security framework for UAV networks,  in detecting and identifying UAVs based on RF signals. Experimental results have demonstrated the effectiveness of the proposed zero-trust security framework in enhancing the security of UAV networks. 
Similarly, Ouiazzane \et \cite{ouiazzanezero} developed a zero-trust security framework for UAV intrusion detection by using network traffic data. 
However, the existing zero-trust security mechanisms are not well integrated with federated learning, which may lead to potential security threats. 

Though there have been extensive works on UAV authentication and detection, the security and privacy of UAV networks are still challenging in three aspects, namely, \emph{task-level}, \emph{data-level}, and \emph{network-level} security. 
\begin{itemize}
    \item Firstly, existing task-level security methods rely on \emph{softmax} classifiers trained for known UAVs, failing to identify \emph{unknown} UAVs (UAV rejection). A robust mechanism for authenticating and rejecting UAVs based on unique features is essential.
    \item Secondly, data-level security requires shifting from centralized deployment to distributed mechanisms that can authenticate and reject UAVs without sharing raw data, reducing single-point failure risks and enhancing network security.
    \item Thirdly, network-level security remains neglected in existing UAV detection methods. A robust security mechanism is needed to protect the confidentiality, integrity, and availability of UAV-transmitted data against potential threats.
\end{itemize}

To advance UAV authentication in wireless networks, we propose a federated learning (FL)-based lightweight spectrogram network (LSNet) with zero trust for UAV authentication and rejection. The proposed model integrates the principles of zero trust security to protect wireless networks from potential security threats.  The FL-based LSNet enables multiple clients to collaboratively train a global deep-learning model without sharing their raw data, thereby preserving the privacy of the data. Moreover, the FL-based LSNet with zero trust ensures that all model submission is authenticated, authorized, and encrypted, thereby minimizing the risk of security breaches. Then, to execute UAV authentication and rejection in a distributed manner, a lightweight spectrogram network with a novel loss function is proposed. The proposed LSNet can effectively authenticate and reject UAVs, thereby enhancing the security of UAV networks. The main contributions of this paper are summarized as follows
\begin{enumerate}
    \item We propose a federated learning-based lightweight network with zero trust for enhancing the security of wireless networks. The framework integrates security at multiple levels simultaneously—task-level (distinguishing known vs. unknown UAVs), data-level (preserving privacy during model training), and network-level (protecting against external threats). This holistic approach provides significantly stronger security guarantees than systems focusing on only one security aspect. The framework enables multiple clients to collaboratively train a global deep-learning model without sharing their raw data, thereby preserving the privacy of the data.
    \item To execute UAV authentication and rejection in a distributed manner, we present a lightweight spectrogram network (LSNet) with a novel loss function. Our experiments reveal that lightweight neural networks ($<$400K parameters) can achieve comparable or superior performance to much larger models for UAV authentication, with LSNet requiring only 1.6MB of storage. Multi-channel attention convolutional (MCAC) blocks effectively capture the spatial and temporal dependencies of spectrogram data. Moreover, our novel class anchor loss minimizes the distance of known UAVs and maximizes the distance of unknown UAVs, enabling successful authentication and rejection.
    \item Extensive experiments demonstrate the effectiveness of our FL-based LSNet, achieving over $80\%$ accuracy for known UAVs and $0.7$ AUROC for unknown UAVs with just 5 clients. Our findings show that full client participation improves accuracy by approximately 9.45\% compared to single-client participation, highlighting the importance of consistent client engagement. Additionally, we observe that UAVs from the same manufacturer (e.g., FutabaT14 and FutabaT7) present classification challenges, indicating areas for future improvement in distinguishing similar UAV types.
\end{enumerate}

The remainder of this paper is organized as follows. 
Section \ref{sec:related} reviews the related works, including UAV authentication and detection mechanisms, federated learning, and zero trust security. 
Section \ref{sec:preliminaries} offers the problem definition. Section \ref{sec:design} describes the algorithm and system design of the proposed federated learning (FL)-based lightweight network with zero trust, including the system model and the lightweight spectrogram network (LSNet) for UAV authentication and rejection. Section \ref{sec:evaluation} presents the simulation results. Section \ref{sec:conclusion} concludes the paper.

\section{Related Works}
\label{sec:related}

In this section, we explore approaches related to UAV authentication and detection tasks. Firstly, we discuss traditional UAV authentication mechanisms. Then, we present the recent UAV authentication mechanisms based on physical signals, such as acoustic, radar, and radio-frequency (RF) signals. Finally, we introduce federated learning and zero-trust security mechanisms. 
Table \ref{tab} summarizes the UAV authentication and detection methods, highlighting their methods, advantages, and limitations.

\subsection{Traditional UAV Authentication Mechanisms} 

\paragraph{PUF-based UAV Authentication} 
PUFs leverage unique hardware characteristics for UAV authentication. Alladi \et \cite{alladi2020secauthuav} proposed a lightweight mutual authentication scheme for UAV-ground station communication, while Zhang \et \cite{zhang2024prlap} developed PRLAP-IoD for securing drone-gateway and drone-drone communications. Though effective for identity verification, PUF-based methods face challenges meeting the lightweight requirements of dynamic UAV networks.

\paragraph{Traffic Data Analysis-based UAV Authentication}
Traffic data analysis approaches use network communication patterns as UAV fingerprints. Bisio \et \cite{bisio2018unauthorized} utilized statistical features from traffic data for UAV authentication, while Alipour \et \cite{alipour2019machine} extracted features from packet size and arrival times in encrypted Wi-Fi communications to enable rapid detection with reduced latency. However, these methods face mounting data privacy concerns \cite{ezuma2019detection} and difficulties accessing proprietary communication channels used by commercial UAVs.

\subsection{Physical Signal-based UAV Authentication Mechanisms}
\begin{table}[h]
\caption{Comparison of UAV Authentication and Detection Methods}
\label{tab}
\centering
\begin{tabular}{|p{1.8cm}|p{1.8cm}|p{3cm}|}
\hline
\textbf{Type} & \textbf{Key Works} & \textbf{Approach \& Limitations} \\
\hline
\multicolumn{3}{|c|}{\textbf{Traditional Authentication Mechanisms}} \\
\hline
PUF-based & \cite{alladi2020secauthuav}, \cite{zhang2024prlap} & Uses physical hardware characteristics for authentication. Limited by complexity of designing lightweight, robust implementations for dynamic UAV networks. \\
\hline
Traffic Data Analysis & \cite{bisio2018unauthorized}, \cite{alipour2019machine} & Extracts features from network traffic for authentication. Challenged by data privacy concerns and proprietary communication channels. \\
\hline
\multicolumn{3}{|c|}{\textbf{Physical Signal-based Methods}} \\
\hline
Acoustic & \cite{anwar2019machine}, \cite{seo2018drone} & Extract MFCC/LPCC features with SVM or CNN classification. Limited by environmental noise sensitivity. \\
\hline
Radar & \cite{messina2019classification}, \cite{zhang2017classification} & Use radar echo processing with FFT or dual-frequency sensing. Limited by line-of-sight requirements. \\
\hline
RF-based (Feature Engineering) & \cite{medaiyese2021machine}, \cite{ezuma2019detection} & Extract features using FFT/STFT with traditional ML models. Limited by the need for expert feature design. \\
\hline
RF-based (Deep Learning) & \cite{xue2023radio}, \cite{ozturk2020rf}, \cite{zhang2023rf} & Apply CNN directly to RF signals or spectrograms with data augmentation. More robust but computationally intensive. \\
\hline
Visual & \cite{thai2019detection}, \cite{rozantsev2016detecting} & Uses camera-based images and videos. Susceptible to environmental conditions like lighting and physical obstructions. \\
\hline
\end{tabular}
\end{table}

\paragraph{Acoustic Signal-based UAV Authentication}
Anwar \et \cite{anwar2019machine} gathered environmental sound signals and extracted features utilizing Mel frequency cepstral coefficients (MFCC) and linear predictive cepstral coefficients (LPCC). Subsequently, a support vector machine (SVM) was employed to classify UAVs. In a similar vein, the authors in \cite{seo2018drone} applied the short-time Fourier transform (STFT) method to convert UAV sound signals into spectrograms, upon which a convolutional neural network (CNN) was utilized to conduct the classification task.

\paragraph{Radar Signal-based UAV Authentication}
The work \cite{messina2019classification} sampled the swept-frequency radar echo and employed feature extraction via a high-pass filter and the Fast Fourier Transform (FFT) to classify unmanned aerial vehicles (UAVs). Similarly, Zhang \et \cite{zhang2017classification} introduced a dual-frequency radar classification approach, in which data were gathered independently through K-band and X-band radar sensors. This was followed by processing with the STFT and ultimately, classification of UAVs was performed using an SVM classifier. 

\paragraph{RF Signal-based Methods}
RF-based UAV authentication has evolved from feature engineering to deep learning approaches. Early methods extracted specific features from RF signals using signal processing techniques. Medaiyese \et \cite{medaiyese2021machine} captured low-frequency spectra from UAV-controller communications as input for XGBoost models, while Ezuma \et \cite{ezuma2019detection} pioneered a Markov-based model with Bayesian decision mechanisms for RF signal detection.

With advances in deep learning, more sophisticated RF-based methods have emerged. Xue \et \cite{xue2023radio} addressed practical challenges like variable operating channels through morphological filtering and data augmentation. Ozturk \et \cite{ozturk2020rf} demonstrated CNNs' enhanced resilience to noise when using both time-series and spectrogram inputs. Zhang \et \cite{zhang2023rf} proposed a low-cost data augmentation method mixing drone signals with background noise to improve robustness.

While these physical signal-based methods successfully detect known UAVs, they typically don't address unknown UAV rejection and rely on centralized implementations that create privacy and single-point failure concerns.

\subsection{Federated Learning}
\label{ssec:fl}
Federated learning (FL) \cite{mcmahan2017communication} solves the problem of data island on the premise of privacy protection. FL has emerged as a promising approach for enhancing the security and privacy of wireless communication networks. FL enables multiple clients to collaboratively train a global learning model without sharing their raw data. This decentralized learning paradigm ensures that the data remains on the local clients, thereby preserving the privacy of the data. However, FL is susceptible to various security threats, such as model poisoning attacks, backdoor attacks, and inference attacks. These attacks can compromise the integrity and confidentiality of the global model, thereby undermining the security and privacy of the FL process \cite{hijazi2023secure}.

\subsection{Zero Trust}
\label{ssec:zt}

Zero trust \cite{stafford2020zero} security is a security model that assumes that all devices, users, and applications are untrusted, regardless of their location. This model ensures that all network traffic is authenticated, authorized, and encrypted, thereby minimizing the risk of security breaches. Zero-trust security mechanisms have been widely used to enhance the security of wireless communication networks \cite{aloqaily2024guest}. The zero-trust security model is based on the principle of least privilege, which restricts access to sensitive data and resources to only authorized users. By implementing zero-trust security mechanisms, wireless communication networks can protect against potential security threats, thereby ensuring the security and privacy of the data transmitted over the network.

\section{Problem Definition}
\label{sec:preliminaries}

\begin{figure}
\centering
\includegraphics[width=0.45\textwidth]{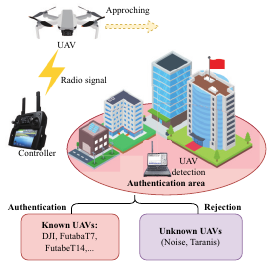}
\caption{UAV authentication and rejection based on radio-frequency signals.}
\label{fig:scenario}
\end{figure}

Fig. \ref{fig:scenario} presents a typical case of UAV authentication and rejection based on radio-frequency (RF) signals. The system is designed to continuously monitor and analyze radio signals, enabling it to promptly detect any UAVs approaching an authentication area. For a known UAV, the system can authenticate the UAV by comparing the received RF signals with the known UAV signals and then predict the UAV type. For an unknown UAV, the system can effectively reject the UAV. 

The UAV authentication and rejection task are treated as a multi-class classification problem with open set recognition. 
Formally, the input features $\mathcal{X}$ of existing UAVs are fed into the network, and then the network predicts the type of the UAVs, i.e., $\mathcal{Y}_i, i=1, 2, \cdots, K$ if the UAV is known, or $y_{U}$ if the UAV is unknown. 
Specifically, the set of UAV type labels can be defined as $\mathcal{Y} = \{y_1, y_2, \cdots, y_K, y_{U}\}$, where $y_i$ denotes the label of the $i$-th UAV type, and $y_{U}$ represents the label of the unknown UAVs. The goal of UAV authentication and rejection is to learn a mapping function $f: \mathcal{X} \rightarrow \mathcal{Y}$.

Given the extensive nature of RF signal data, the proposed framework is tasked with distinguishing pertinent signals amidst the inherent noise and disruptions present in a real-world wireless environment, including interferences from concurrent radio users such as WiFi and Bluetooth devices. Consequently, the feature set $\mathcal{X}$ is derived from a comprehensive collection of RF signals $r(t)$ associated with UAV types, followed by subsequent signal processing efforts. In a generalized form, the received RF signal $r(t)$ is defined by
\begin{equation}
r(t) = \phi(e(t)) + n(t),
\end{equation}
where $e(t)$ represents the signal emitted by the $k$-th UAV controller, and $n(t)$ signifies the additive white Gaussian noise (AWGN) characterized by zero mean and variance $\sigma^2$. The function $\phi(\cdot)$ encapsulates the channel response, influenced by variables such as the distance between the UAV and the receiver, antenna gain, and path loss. Following reception, the signal $r(t)$ undergoes processing to extract the features $\mathcal{X}$, which are essential for the UAV authentication and rejection procedures. Advanced signal processing techniques such as the discrete Fourier transform (DFT), short-time Fourier transform (STFT), and Hilbert-Huang transform (HHT) are employed to derive effective features that are compatible with and informative for the predictive models.

\section{Algorithm and System Design}
\label{sec:design}
This section presents the algorithm and system design of the proposed UAV authentication and rejection. 
We first introduce the system model and the lightweight spectrogram network (LSNet) for UAV authentication and rejection. 
Then, we present the loss function for UAV authentication and rejection.

\subsection{System Model}
\label{ssec:model}

\begin{figure}
\centering
\includegraphics[width=0.48\textwidth]{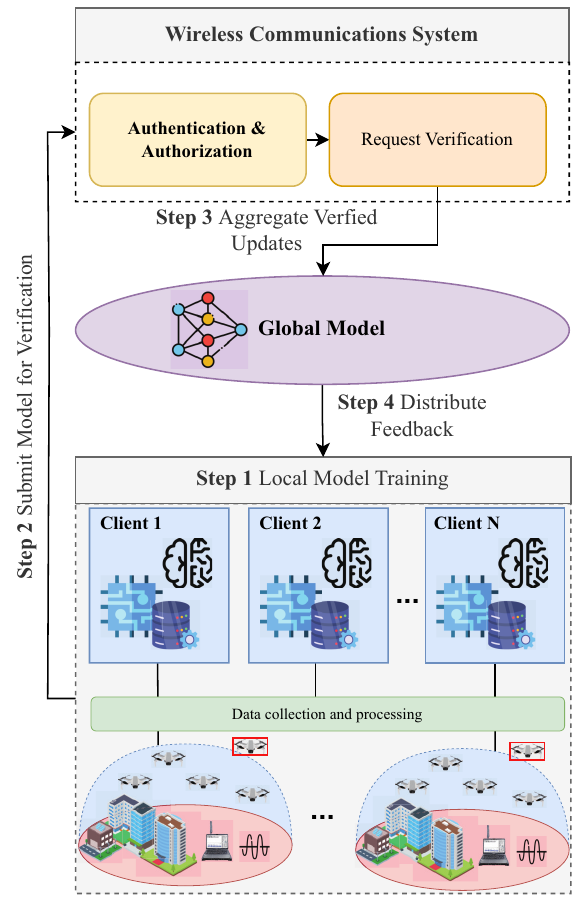}
\caption{The architecture of the proposed federated learning (FL)-based lightweight network with zero trust for UAV authentication and rejection.}
\label{fig:fl}
\end{figure}

As shown in Fig. \ref{fig:fl}, we propose a federated learning (FL)-based lightweight network with zero trust for UAV authentication and rejection, inspired by the work \cite{asad2024integrative}. The FL framework consists of four main steps, 1) local model training, 2) model submission and verification, 3) global model aggregation, and 4) global model deployment. The function of each step is detailed as follows.
\begin{enumerate}
    \item \textbf{Local Model Training:}  In each client, data are collected and then preprocessed as spectrograms. These spectrograms are then fed into the local model for training as the local dataset by using the local loss function. 
    \item \textbf{Model Submission and Verification:} After the local model training, the local model is submitted to the server for verification. The server verifies the submitted local models by using zero-trust security mechanisms to ensure that the models are authentic and authorized. 
    \item \textbf{Global Model Aggregation:} The server aggregates the verified local models using the federated average algorithm to obtain a new global model. The global model is then deployed to all clients for the next round of training.
    \item \textbf{Global Model Deployment:} The global model is deployed to all clients for the next round of training. The clients use the global model to update their local models and then submit the updated local models to the server for verification. The server verifies the updated local models and aggregates them to obtain a new global model. The global model is then deployed to all clients for the next round of training.
\end{enumerate}

To better understand the FL-based framework with zero trust, we present the algorithm of the proposed FL-based LSNet with zero trust in Algorithm \ref{alg:training}. The function of each step is detailed as follows. 

\begin{algorithm} 
\caption{The algorithm of the proposed FL-based lightweight network with zero trust.} 
\label{alg:training} 
\begin{algorithmic}[1] 
\REQUIRE spectrogram data $\mathcal{X}$, labels $\mathcal{Y}$ for $K$ clients, the number of rounds $T$, the batch size $B$, the learning rate $\eta$;
\ENSURE The global model $\bm{\omega_T}$;
\STATE \textbf{Server executes:}
\STATE \quad Initialize the global model $\bm{\omega_0}$;
\STATE \quad \textbf{for} round $t=0,1,2,\ldots$, T \textbf{do}
\STATE \quad \quad Select a subset of clients $\mathcal{K}_t$;
\STATE \quad \quad \textbf{for} each client $k \in \mathcal{K}_t$ \textbf{do}
\STATE \quad \quad \quad Send the global model $\bm{\omega_t}$ to client $k$;
\STATE \quad \quad \quad Client $k$ updates the local model $\bm{\omega_t}^k$ by minimizing the local loss $\mathcal{L}_k(\bm{\omega_t}^k)$;
\STATE \quad \quad \quad Send the updated local model $\bm{\omega_t}^k$ to the server;
\STATE \quad \quad \quad Verify the updated local model $\bm{\omega_t}^k$;
\STATE \quad \quad \textbf{end for}
\STATE \quad \quad Aggregate the updated local models to obtain the new global model $\bm{\omega_{t+1}}$;
\STATE \quad \quad Broadcast the new global model $\bm{\omega_{t+1}}$ to clients;
\STATE \quad \textbf{end for}
\STATE \textbf{return }The global model $\bm{\omega_T}$.
\end{algorithmic}
\end{algorithm}

\subsection{Lightweight Spectrogram Network (LSNet)}
\begin{figure*}
\centering
\includegraphics[width=0.95\textwidth]{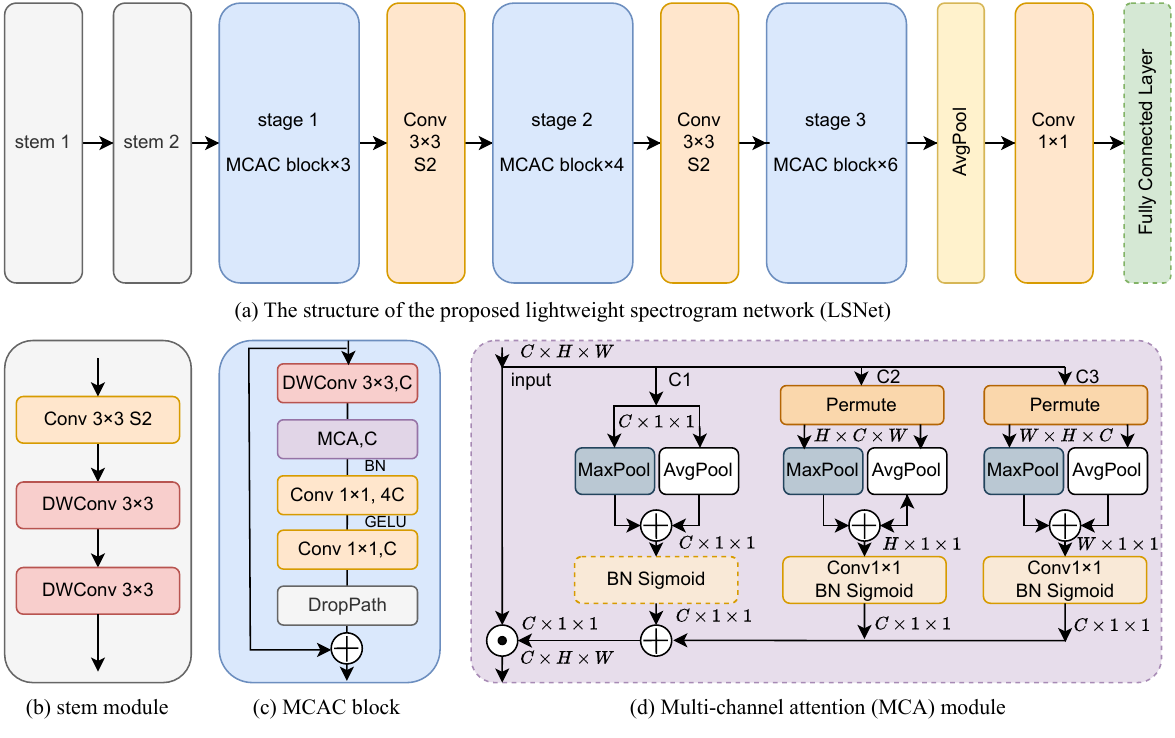}
\caption{The architecture of the proposed lightweight spectrogram network (LSNet) for UAV authentication and rejection. }
\label{fig:model}
\end{figure*}

Fig. \ref{fig:model} illustrates the architecture of the proposed lightweight spectrogram network (LSNet), designed for efficient UAV recognition. The network comprises several stages, each with different components tailored for varying levels of feature extraction and refinement. In Fig. \ref{fig:model} (a), the network begins with two initial stem modules which are designed to preprocess the input data. The stem modules consist of a series of depthwise separable convolutions followed by a pointwise convolution, aiming to reduce computational cost while maintaining effectiveness in capturing features. 
After the initial stem modules, the network proceeds to a series of multi-channel attention convolution (MCAC) blocks, as shown in Fig.\ref{fig:model} (b). The MCAC blocks are designed to capture and refine features across multiple channels, enhancing the network's ability to focus on important features in the input data. Before each stage of MCAC blocks, a convolutional layer with a stride of $2$ is utilized to gradually reduce spatial dimensions while increasing the depth of features. 
The network concludes with a global average pooling (GAP) layer to aggregate features across spatial dimensions, followed by a $1\times 1$ convolution to consolidate the learned features into a final output representation. 
Finally, a fully connected (FC) layer is adopted to generate the final output. The detailed architecture of the LSNet is presented in Table \ref{tab:lsnet}. 
The mentioned components are elaborated in the following subsections. 

\begin{table}[]
\centering
\caption{The detailed architecture of the proposed lightweight spectrogram network (LSNet).}
\label{tab:lsnet}
\begin{tabular}{ccl}
\toprule
layer&structure& output size \\
\midrule
stem 1                      & 
\begin{tabular}[c]{@{}c@{}}Conv$3\times 3$, 16, stride=2\\ DWConv$3\times 3$, 16\\ DWConv$3\times 3$, 16\end{tabular}        & $16\times 64\times 64 $   \\
\midrule
stem 2                      & 
\begin{tabular}[c]{@{}c@{}}Conv$3\times 3$, 16, stride=2\\ DWConv$3\times 3$, 16\\ DWConv$3\times 3$, 16\end{tabular}        
& $16\times 32\times 32 $    \\
\midrule
stage 1& 
$\left [\begin{matrix}
    \text{DWConv}3\times 3,16 \\
    \text{MCA},16\\
    \text{Conv}1\times 1,64\\
    \text{Conv}1\times 1,16
\end{matrix}\right ]\times 3$
& $16\times 32\times 32 $   \\
\midrule
conv& Conv$3\times 3$, 32, stride=2& $32\times 16\times 16 $     \\
\midrule
stage 2& 
$\left [\begin{matrix}
    \text{DWConv}3\times 3,32 \\
    \text{MCA},32\\
    \text{Conv}1\times 1,128\\
    \text{Conv}1\times 1,32
\end{matrix}\right ]\times 4$
& $32\times 16\times 16 $    \\
\midrule
conv& Conv$3\times 3$, 64, stride=2& $64\times 8\times 8 $\\
\midrule
stage 3                     & 
$\left [\begin{matrix}
    \text{DWConv}3\times 3,64 \\
    \text{MCA},64\\
    \text{Conv}1\times 1,256\\
    \text{Conv}1\times 1,64
\end{matrix}\right ]\times 6$
& $64\times 8\times 8 $      \\
\midrule
GAP    & AvgPool($1\times 1$) & $64\times 1\times 1 $\\
\midrule
conv & Conv$1\times 1$, 128& $128$\\
\midrule
fc     & FC(128, 7) & $7$ \\
\midrule
params & $358,023$ & 1.6M \\
\bottomrule
\end{tabular}
\end{table}

\subsubsection{Stem Module}

ResNet \cite{he2016deep} employs a $7\times 7$ convolution with a downsampling rate of $2\times$ followed by a $3\times 3$ max pooling with a stride of $2$, achieving a fourfold reduction in spatial dimensions. This configuration leads to overlapping sampling regions, which enhances the redundancy in the perceived features. In contrast, the ConvNeXt model \cite{liu2022convnet} utilizes a $4\times 4$ convolution with a stride of $4$, employing a non-overlapping downsampling technique. This approach may result in a potential loss of detailed features and fine-grained information. However, the $3\times 3$ convolution proposed in the VGGNet \cite{simonyan2014very} model is optimal for extracting detailed depth features. To effectively capture a rich array of redundant fine-grained features, the stem module is strategically designed. Initially, a $3\times 3$ standard convolution with a stride of $2$ processes the input features.  Subsequently, this is followed by two $3\times 3$ depthwise separable convolutions with a stride of $1$, which not only increase the receptive field but also enhance the extraction of local fine-grained features. Each depthwise separable convolution is followed by a group normalization and a GELU activation function, as recommended by Hendrycks and Gimpel \cite{hendrycks2016gaussian}, to improve the model's accuracy.

\subsubsection{Multi-Channel Attention Convolutional (MCAC) Block}
As shown in Fig. \ref{fig:model} (c), the Multi-Channel Attention Convolutional (MCAC) block is a key component of the LSNet, designed to enhance the feature representation for RF spectrograms. 
The proposed MCAC block consists of a depthwise convolutional layer, a multi-channel attention (MCA) module, two $1\times 1$ convolutional layers, and a Droppath layer. 
The depthwise convolutional layer with $3\times 3$ kernel and stride $1$ is used to extract fine-grained features from the input features. 
The MCA module is designed to learn the features from three dimensions, i.e., channel, height, and width, respectively, and fuse them to generate the MCA attention. The MCA attention is then multiplied with the input tensor to enhance the feature representation. 
After the MCA module, a batch normalization layer is applied to normalize the feature map, followed by a GELU activation function to introduce non-linearity. 
Then, the first $1\times 1$ convolutional layer is used to increase the channel dimension of the feature map by a factor of four, followed by a GELU activation function. 
The second $1\times 1$ convolutional layer is used to decrease the channel dimension of the feature map to the original size. 
After that, the Droppath \cite{larsson2016fractalnet} layer is applied to prevent overfitting by randomly dropping the connections in the network. 
Finally, the residual connection is added to the output of the MCAC block to facilitate the flow of information through the network. The MCAC block is designed to capture and refine features across multiple channels, enhancing the network's ability to focus on important features in the input data. The above operations can be formulated as
\begin{equation}
    x_o = x_i + \text{DPath}(\text{Conv}(\text{GELU}(\text{Conv}(\text{BN}(\text{MCA}(\text{DW}(x))))))),
\end{equation}
where $x_i$ and $x_o$ denote the input and output of the MCAC block, respectively. BN denotes the batch normalization layer, DW denotes the depthwise convolutional layer, MCA denotes the multi-channel attention module, Conv denotes the convolutional layer, GELU denotes the GELU activation function, and DPath denotes the Droppath layer. 

\subsubsection{Multi-Channel Attention (MCA) Module}
Before each stage of MCAC blocks, a spatial downsampling layer is applied to reduce the spatial dimensions while increasing the depth of features. 
Prior to each stage of the MCAC blocks, a spatial downsampling layer is incorporated to reduce the spatial dimensions while concurrently enhancing the depth of the feature representations. The downsampling layer comprises a $3\times 3$ convolution with a stride of 2, followed by a batch normalization layer. This methodology is consistent with the approaches found in established models such as ResNet \cite{he2016deep}, ConvNeXt \cite{liu2022convnet}, and SwinTransform \cite{liu2021swin}, where spatial downsampling is facilitated through the deployment of separate layers, integrating a normalization layer and a $2\times 2$ convolution with a stride of 2. In our implementation, similar spatial downsampling techniques are employed, utilizing a normalization layer alongside a $3\times 3$ convolution layer with a stride of 2, positioned between the stages to effectively manage the feature scale transitions.

The Multi-Channel Attention (MCA) module, as depicted in Fig. \ref{fig:model} (d), is a sophisticated component designed to enhance the feature representation in convolutional neural networks by focusing on salient features across different channels. This selective emphasis improves the model's performance in tasks involving complex feature extractions. The MCA module consists of four parallel branches, each performing a unique operation to generate attention maps. 
\begin{itemize}
    \item The first branch is the residual connection, which directly passes the input to the output.
    \item The second branch $C1$ first computes max pooling and average pooling across the channel dimension, followed by an addition operation. Then, a batch normalization layer and a sigmoid activation function are applied to generate the attention map.
    \item The third branch $C2$ first permutes the input tensor $C\times H\times W$ into $H\times C\times W$, then applies max pooling and average pooling across the dimension. Then, an addition operation is added, generating an attention map with $H\times 1\times 1$. Finally, the output is passed into a $1\times 1$ convolutional layer with a batch normalization layer and a sigmoid activation function to generate the final attention map with $C\times 1\times 1$.
    \item The fourth branch $C3$ first permutes the input tensor $C\times H\times W$ into $W\times H\times C$, then applies max pooling and average pooling across the dimension. Then, an addition operation is added, generating an attention map with $W\times 1\times 1$. Finally, the output is passed into a $1\times 1$ convolutional layer with a batch normalization layer and a sigmoid activation function to generate the final attention map with $C\times 1\times 1$.
\end{itemize}
The enhanced feature representations from the last three branches are added together and then multiplied with the input tensor to generate the final output of the MCA module. This dynamic and adaptive mechanism allows the model to focus on the most informative parts of the input data, enhancing the model's discriminative power and efficiency in high-stakes applications.

\subsection{Loss function for UAV Authentication and Rejection}
To authenticate and reject UAVs simultaneously, we propose a more general loss for the UAV authentication and rejection task. 
The authentication and rejection task is treated as a multi-class classification problem, where the known UAVs are classified into $N$ classes, and the unknown UAVs are classified into the $N+1$ class. 
During the training process, the data of unknown UAVs can not be accessed by the clients. 
Thus, we introduce the concept of \emph{class centers} to represent the known UAVs. 
The proposed LSNet $f_{LS}$ is used for extracting features from the input RF signals, projecting an input spectrogram $x$ to a vector of class logits $z=f_{LS}(x)$. 
Then, for each known UAV class $i$, we define a class center $\mathbf{c}_i$ in the logit space, where $i=1, 2, \cdots, N$. 
Given an N-dimensional logit space corresponding to $N$ known classes, we position the anchored center for each class along its respective class coordinate axis. This placement means that the anchored center point is effectively equivalent to a scaled standard basis vector $\mathbf{e}_i$ or a scaled one-hot vector for each class. The scale factor $\alpha$, which determines the magnitude of the anchored center, is a hyperparameter. This process can be given as
\begin{equation}
    \mathbf{C}=(\mathbf{c}_1, \cdots, \mathbf{c}_N) = (\alpha \cdot \mathbf{e}_1, \cdots, \alpha \cdot \mathbf{e}_N),
\end{equation}
\begin{equation}
    \mathbf{e}_1=(1, 0, \cdots, 0)^T, \cdots, \mathbf{e}_N=(0, 0, \cdots, 1)^T.
\end{equation}

A new layer, $e(\mathbf{z}, \mathbf{C})$, that calculates $\mathbf{d}$, a vector of Euclidean distances between a logit vector $\mathbf{z}$ and the set of class centers $C$. Thus, the output of the distance-based classifier is
\begin{equation}
    \mathbf{d} = e(\mathbf{z}, \mathbf{C}) = (||\mathbf{z} - \mathbf{c}_1||_2, ||\mathbf{z} - \mathbf{c}_2||_2, \cdots, ||\mathbf{z} - \mathbf{c}_N||_2)^T,
\end{equation}
where $||\cdot||_2$ denotes the Euclidean norm.

In the training phase, the objective is to develop an embedding $f_{LS}(x)$ within the logit space where inputs associated with known UAVs form compact clusters specific to each class. Such clustering facilitates the application of a distance-to-class-center metric in the testing phase. This metric is crucial for rejecting inputs from unknown UAVs and accurately classifying inputs from known UAVs. To achieve this, it is essential to implement a distance-based loss function that accomplishes two goals, a) minimizing the distance between training inputs and their corresponding class center, and b) maximizing the distance between these inputs and the centers of all other UAVs to promote discriminative learning.

To achieve the aforementioned objectives, we employ a modified tuplet loss term, denoted as $\mathcal{L}_T$ \cite{sohn2016improved}, which compels an input $x$ to maximize the discrepancy between its distance to the correct class center $\mathbf{c}_y$ and its distances to all other class centers. The loss term is defined as
\begin{equation}
    \mathcal{L}_T(x,y) = \log(1 + \sum_{j\neq y}^N \exp(d_{y} - d_{j})),
\end{equation}
where $\mathcal{L}_T$ diverges from the conventional Tuplet loss \cite{sohn2016improved} by focusing on class centers $\mathbf{C}$ instead of sampled class instances. This modification equates the Tuplet loss term to a cross-entropy loss that is applied to the distance vector $\mathbf{d}$. However, instead of employing a \texttt{softmax} function, a \texttt{softmin} function is utilized. The \texttt{softmin} function serves as the inverse of \texttt{softmax}, and it allocates a high value (approximately 1) to the minimal value within the input vector. The \texttt{softmin} function is defined by
\begin{equation}
    \text{softmin}(\mathbf{d}) = -\log\left(\frac{\exp(-\mathbf{d})}{\sum_{j=1}^N \exp(-\mathbf{d}_j)}\right).
\end{equation}

Although $\mathcal{L}_T$ is effective for discriminative learning by maximizing the margin between the distance to the correct class center and the distances to the incorrect class centers, it does not explicitly ensure that the input minimizes its absolute distance to the correct class center. To address this limitation, we also incorporate a penalty for the Euclidean distance between the training logit and the ground truth class center. This additional penalty is termed the Anchor loss term, which directly encourages inputs to reduce their absolute distance to the correct class center, enhancing the alignment with the desired class-specific clustering.
\begin{equation}
    \mathcal{L}_A(x,y) =\mathbf{d}_y= ||f_{LS}(x) - \mathbf{c}_y||_2.
\end{equation}

\begin{figure}
\centering
\includegraphics[width=0.48\textwidth]{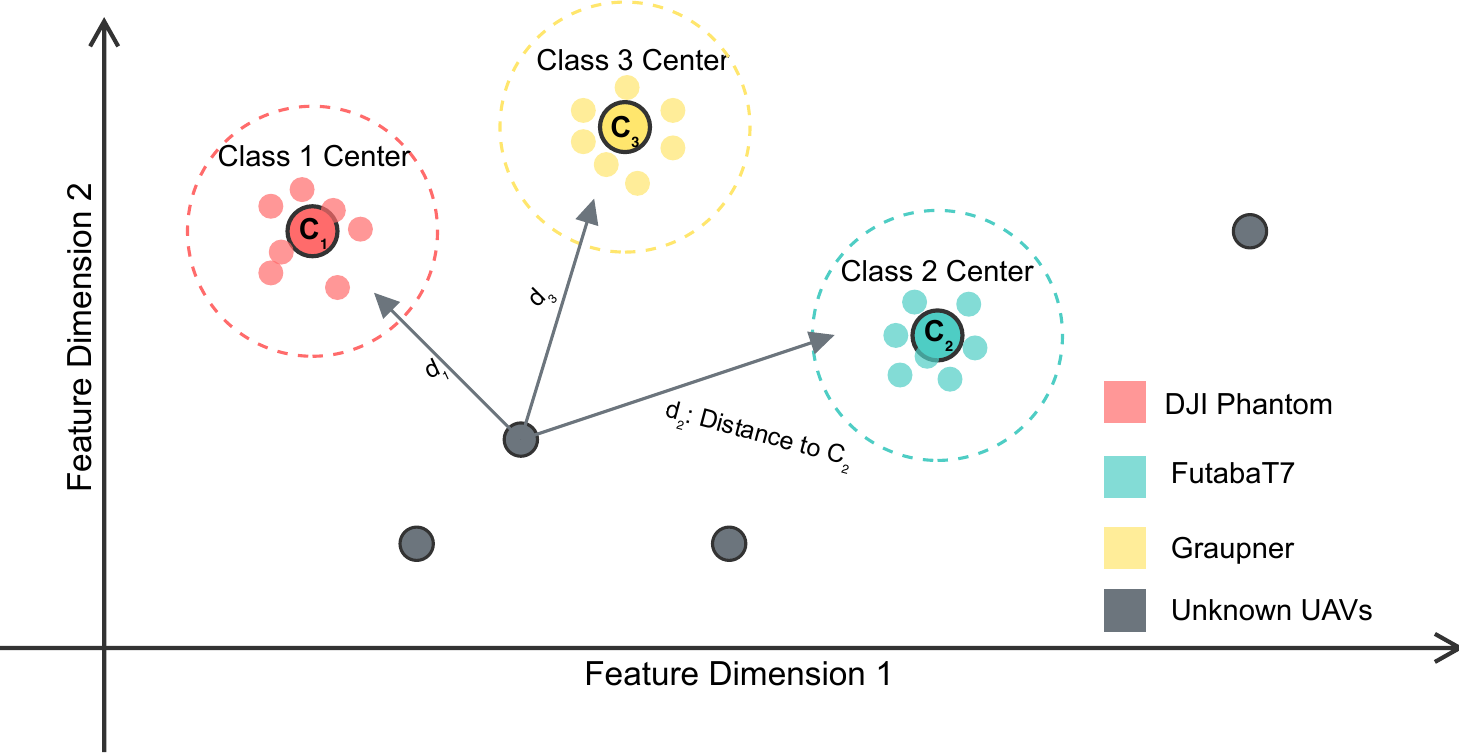}
\caption{Visualization of the Class Anchor loss function in feature space.}
\label{fig:loss_vis}
\end{figure}

By integrating the anchor and tuplet loss terms, we formulate our composite distance-based loss, which we designate as the class anchor (CA) loss, given by 
\begin{equation}
    \mathcal{L}_{CA}(x,y) = \mathcal{L}_T(x,y) + \lambda \mathcal{L}_A(x,y),
\end{equation}
where $\lambda$ is a hyperparameter that balances the two individual loss terms. The integration of the anchor and tuplet loss terms in our loss function serves to minimize the distance of training inputs to their ground-truth anchored class center while simultaneously maximizing the distance to other anchored class centers. Fig. \ref{fig:loss_vis} illustrates the visualization of the class anchor loss function in the feature space. The CA loss function effectively encourages the model to learn a discriminative feature representation that can accurately classify known UAVs while rejecting unknown UAVs.

\section{Experiments and Results}
\label{sec:evaluation}

\subsection{Evaluation Metrics}
UAV authentication and rejection are evaluated using the following metrics: accuracy for known UAVs, and the Area Under the Receiver Operating Characteristic (AUROC) curve for unknown UAVs. 
The accuracy for known UAVs is defined as the ratio of the number of correctly classified known UAVs to the total number of known UAVs. 
The AUROC curve for unknown UAVs is used to evaluate the performance of the model in distinguishing unknown UAVs from known UAVs. 
The AUROC curve is calculated by plotting the true positive rate (TPR) against the false positive rate (FPR) at various threshold settings. 
The AUROC value ranges from $0$ to $1$, where a value of $0.5$ indicates a model with no discrimination ability, equivalent to random guessing. 

Moreover, Open set recognition is a challenging problem in DL-based recognition tasks. 
To better understand the difficulty of the open set recognition problem, the \emph{openness}  \cite{scheirer2012toward} metric is defined as 
\begin{equation}
	openness = 1 - \sqrt{\frac{2\times N_{tr}}{N_{tr}+N_{te}}},
\end{equation}
where $N_{tr}$ represents the number of known modulation schemes in the training set, and $N_{te}$ is the number of known modulation schemes in the testing set. Typically, the openness value is between 0 and 1. A higher openness value indicates a more challenging open set recognition problem. 
$N_{tr} \leq N_{te}$. When $N_{tr} = N_{te}$, the value of $openness$ is 0 and it is closed set recognition.

\subsection{Experimental Setup}

Experiments are conducted on a workstation with an AMD Ryzen 9 7900X3D CPU, 64GB RAM, and an NVIDIA GeForce RTX 4070 Ti GPU, and the system is running Ubuntu 22.04. 
Code is implemented in Python and the PyTorch deep learning framework \cite{paszke2019pytorch}. 
To train our proposed LSNet in a center manner, we employ the SGD \cite{ruder2016overview} optimization algorithm with an initial learning rate of $0.01$ over $50$ training epochs as in \cite{zhang2024sswsrnet} and \cite{zhang2025fsos}. 
For training the LSNet under a federated learning manner, we use the SGD optimization algorithm with an initial learning rate of $0.05$ over $20$ training rounds for each local client. 
The batch size is set to $64$ for both the centralized and federated learning settings. 
The hyperparameters $\alpha$ and $\lambda$ are set to $0.1$ and $0.1$, respectively.

\subsection{Datsets}

In this work, we use the Noisy DroneRF dataset \cite{gluge2023robust} for training and testing the proposed model, which is newly published and contains a large number of RF signals from various UAVs. 
These signals were received by a LogPer antenna and sampled and stored by an Ettus Research USRP B210. 
All signals were recorded at a sampling frequency of 56 MHz (highest possible real-time bandwidth). 
To reduce memory consumption and computational effort, we reduced the bandwidth of the signals by downsampling from 56 MHz to 14 MHz using the SciPy \cite{virtanen2020scipy} \texttt{signal.decimate} function with an 8th order Chebyshev type I filter.

\begin{figure}
\centering
\subfigure[]{
\includegraphics[width=0.95\linewidth]{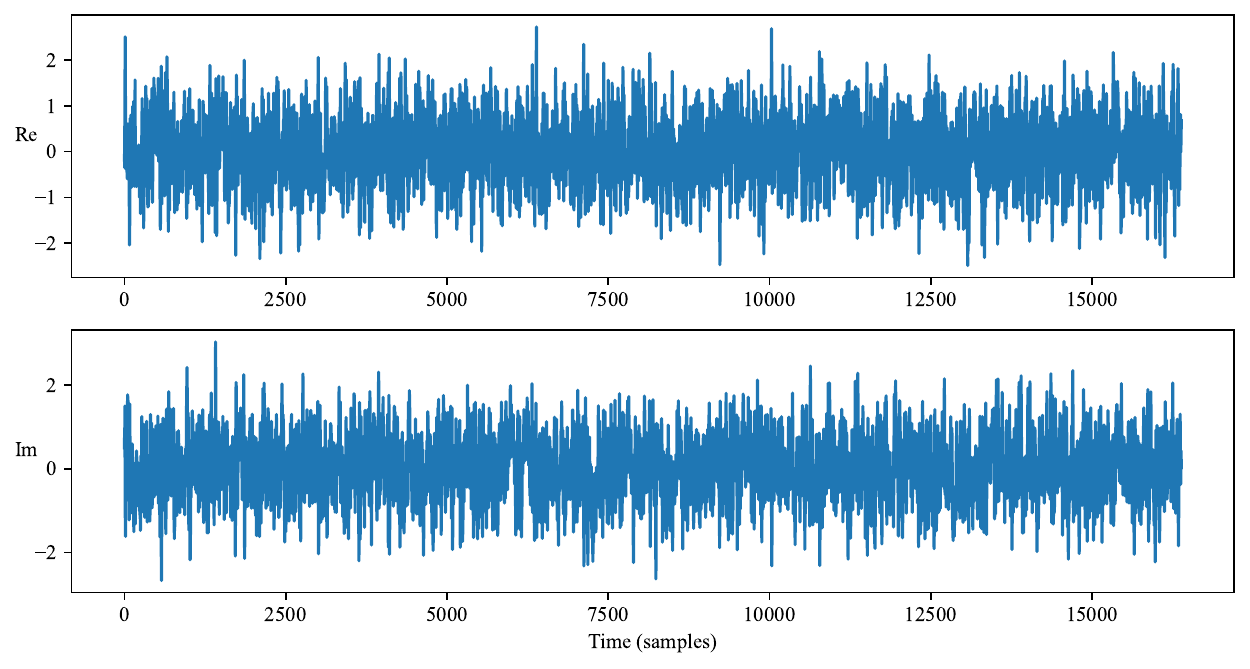}}
\subfigure[]{\includegraphics[width=0.95\linewidth]{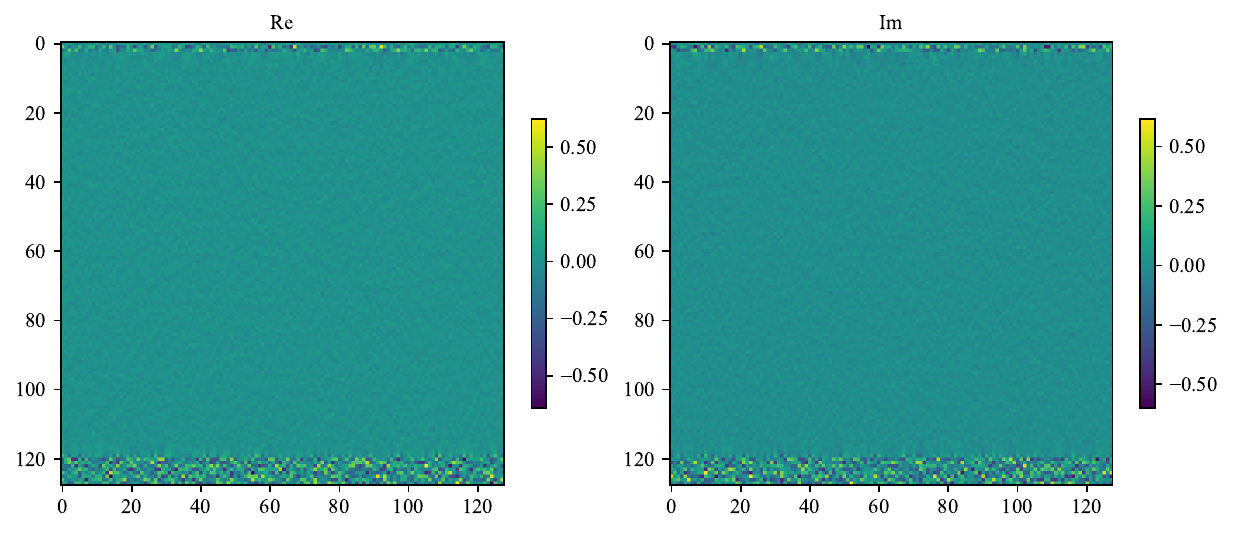}}
\DeclareGraphicsExtensions.
\caption{DJI Drone RF signal at 10dB with duty cycle 1.0. (a) IQ signal. (b) spectrogram.}
\label{fig:data}
\end{figure}

The dataset includes non-overlapping signal vectors of length $16,384$, which corresponds to about $1.2$ ms at $14 $MHz, with added noise sources such as Bluetooth, Wi-Fi, and Gaussian noise. It is structured to challenge algorithms with a realistic mix of drone and noise signals across various signal-to-noise ratios (SNR), ranging from $-20$ to $30$ dB in steps of $2$ dB, \emph{i.e.} $3792-3800$ samples per SNR level. 
Table \ref{tab:dataset} provides an overview of the transmitters and receivers recorded in the dataset and their respective labels. Fig. \ref{fig:data} shows an example of a DJI drone RF signal at $10$ dB with a duty cycle of $1.0$. 

\begin{table*}[]
\caption{Transmitters and receivers recorded in the dataset and their respective labels.}
\centering
\begin{tabular}{cccccccc}
\toprule
Transmitter           & Receiver          & Center Freq. (GHz) & Spacing (MHz) & Duration (ms) & Repetition (ms)   & \textbf{Label}     & \textbf{\#samples} \\
\midrule
DJI Phantom GI 300F   & DJI Phantom 4 Pro & 2.44175      & 1.7     & 2.18     & 630          & DJI       & 2194                          \\
FutabaT7C             & -                 & 2.44175      & 2       & 1.7      & 288          & FutabaT7  & 3661                          \\
FutabaT14SG           & Futaba R7008SB    & 2.44175      & 3.1     & 1.4      & 330          & FutabeT14 & 6938                          \\
Graupner              & Graupner GR-16    & 2.44175      & 1       & 1.9/3.7  & 750          & Graupner  & 6481                          \\
Bluetooth/Wi-Fi Noise & -                 & 2.44175      & -       & -        & -            & Noise     & 52552                         \\
Taranis               & X8R Receiver      & 2.440        & 1.5     & 3.1/4.4  & 420          & Taranis   & 16546                         \\
Turmigy               & -                 & 2.445        & 2       & 1.3      & 61, 120-2900 & Turnigy   & 10333                         \\
\bottomrule
\end{tabular}
\label{tab:dataset}
\end{table*}

The dataset is divided into training and testing sets, with $80\%$ of the data used for training and $20\%$ for testing. 
The dataset is highly imbalanced, with each UAV having a different number of samples. Thus, for federated learning, we use a balanced dataset by selecting the same number of samples for each UAV, which is $2,000$.

\subsection{Performance Comparison}

\begin{figure}
    \centering
    \includegraphics[width=0.47\textwidth]{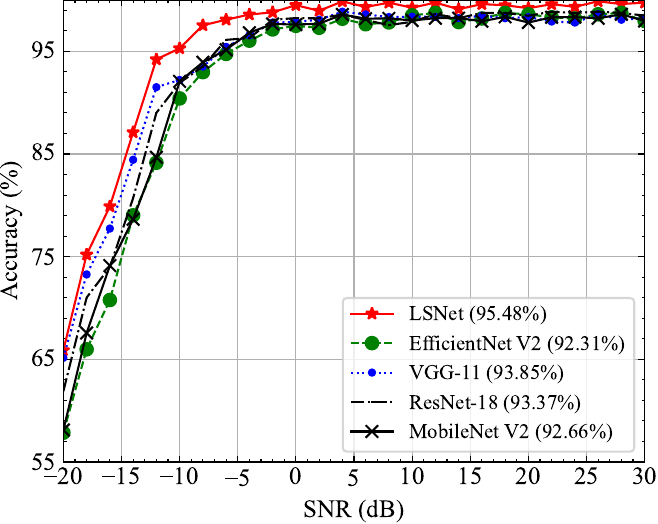}
    \caption{Performance comparison of the proposed LSNet with benchmark models. }
    \label{fig:comparison}
\end{figure}

To evaluate the performance of the proposed LSNet, we compare it with the state-of-the-art methods, including the benchmark VGGNet \cite{simonyan2014very}, ResNet \cite{he2016deep}, MobileNet V2 \cite{sandler2018mobilenetv2}, and Efficientnet V2 \cite{tan2021efficientnetv2}. 
We train these models by using the whole dataset in a centralized manner. 
As shown in Fig. \ref{fig:comparison}, the proposed LSNet outperforms the benchmark models in terms of accuracy, achieving an overall accuracy of $95.48\%$. 
Specifically, the LSNet achieves an accuracy of over $95\%$ for UAV detection when the signal-to-noise ratio (SNR) is larger than $-10$ dB. 
The well-known benchmark VGG11 model achieves an accuracy of $93.85\%$, while the ResNet18 model achieves an accuracy of $93.37\%$. 
This is mainly because these two models have excessive parameters and are prone to overfitting. 
The MobileNet V2 model achieves an accuracy of $92.31\%$, while the EfficientNet V2 model achieves an accuracy of $92.66\%$. 
These two lightweight models have fewer parameters, but they are not as effective as the proposed LSNet, especially when the SNR is low.

A notable observation from Fig. \ref{fig:comparison} is the performance degradation of all models at lower SNR levels, particularly in the range of $-20$ to $-10$ dB. In this challenging noise environment, LSNet demonstrates superior robustness compared to the benchmark models. At $-15$ dB, LSNet achieves approximately $87\%$ accuracy, while other models' performance drops to around $75$-$80\%$. This enhanced performance can be attributed to LSNet's specialized architecture—specifically the multi-channel attention convolutional (MCAC) blocks—which are designed to effectively extract discriminative features from RF spectrograms even when heavily corrupted by noise. The MCAC blocks help focus the network's attention on the most informative parts of the input signal, allowing better feature representation in high-noise conditions. 
In contrast, the general-purpose architectures of the benchmark models, while effective at higher SNR levels, lack the specialized feature extraction mechanisms necessary for RF signal classification in challenging noise environments. VGG11 and ResNet18, despite their deeper architectures, suffer from their excessive parameterization, which can lead to overfitting when signal characteristics become less distinct at lower SNR levels. Similarly, while MobileNet V2 and EfficientNet V2 are designed to be parameter-efficient, they are not optimized for the specific characteristics of RF spectrograms, resulting in their inferior performance particularly below $-10$ dB SNR.

\begin{figure}
    \centering
    \includegraphics[width=0.47\textwidth]{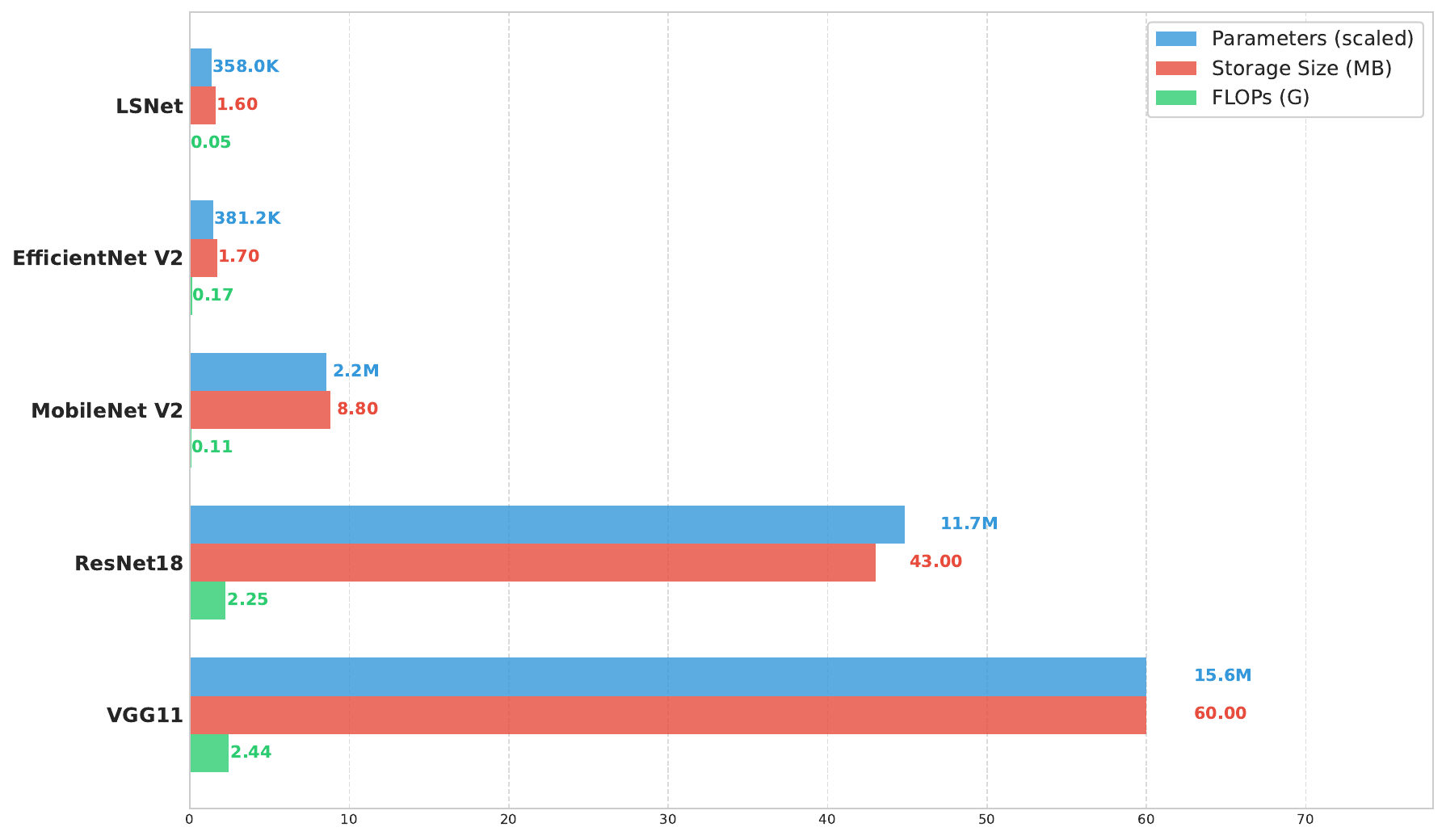}
    \caption{\color{blue}The number of parameters, storage size (MB), and FLOPs of the proposed LSNet and benchmark models.}
    \label{fig:params}
\end{figure}

To further demonstrate the efficiency of the proposed LSNet, we compare the number of parameters, the storage size, and the computational complexity (measured in FLOPs) of the LSNet with the benchmark models, as presented in Fig. \ref{fig:params}. 
It can be observed that the LSNet has fewer parameters and a smaller storage size compared to the benchmark models, with only $358,023$ trainable parameters and a storage size of $1.6$ MB. 
More importantly, LSNet demonstrates superior computational efficiency with only $0.05$G FLOPs, which is $2.2$ times fewer than EfficientNet V2 ($0.17$G), $2.2$ times fewer than MobileNet V2 ($0.11$G), $45$ times fewer than ResNet18 ($2.25$G), and $49$ times fewer than VGG11 ($2.44$G). This significant reduction in computational complexity makes LSNet particularly suitable for deployment in resource-constrained UAV systems. The EfficientNet V2 model has a similar number of parameters and storage size as the LSNet, but it requires more than three times the computational operations and is less effective in terms of accuracy. As for another lightweight model, the MobileNet V2 model has more parameters and a larger storage size than the LSNet, with over $6$ times the number of parameters, over $5$ times the storage size, and more than twice the computational complexity. Though ResNet18 and VGG11 perform well in terms of accuracy, they have a large number of parameters, a large storage size, and significantly higher computational demands, which is not appropriate for resource-constrained devices in UAV networks where both model size and computational efficiency are critical factors.

\begin{figure}
\centering
\subfigure[]{
\includegraphics[width=0.95\linewidth]{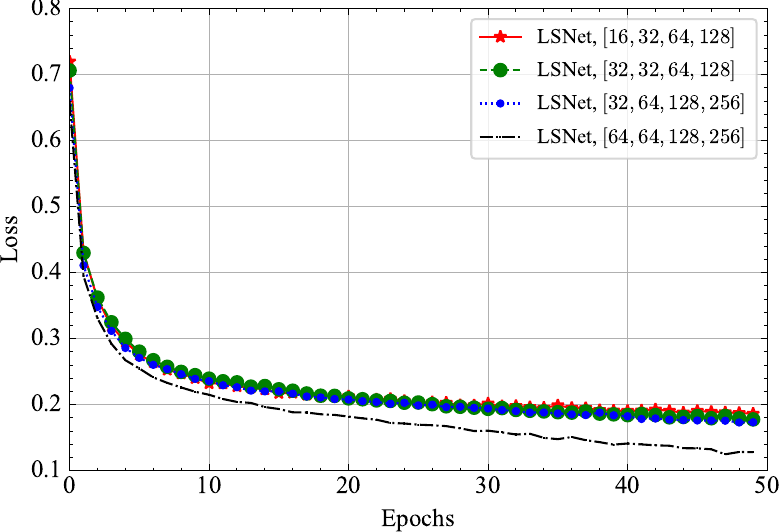}}
\subfigure[]{\includegraphics[width=0.95\linewidth]{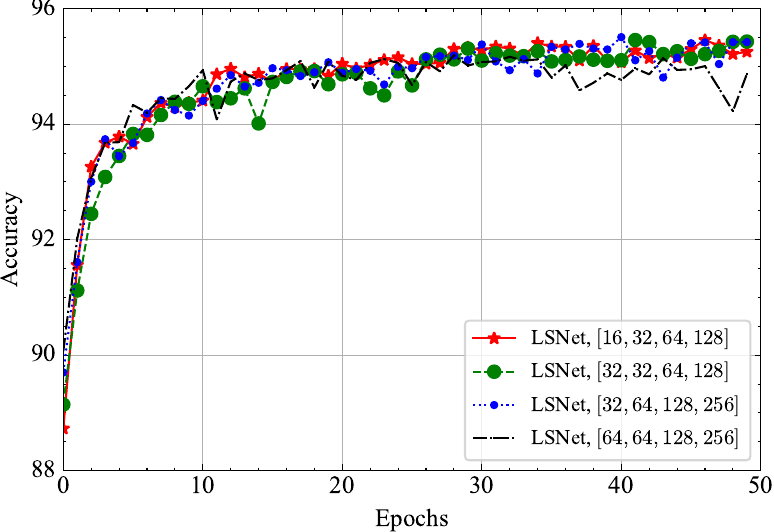}}
\DeclareGraphicsExtensions.
\caption{The training process of LSNet with different channels (a) train loss; (b) test accuracy.}
\label{fig:loss}
\end{figure}

\subsection{Ablation Study}

We follow the ResNet-18 architecture and adopt $[3, 4, 6]$ for the three-stage architectures in the LSNet. 
Thus, to investigate the impact of the channel number in the LSNet, we conduct an ablation study by varying the number of channels in the LSNet. 
We first use a channel number of $[16, 32, 64, 128]$ in the LSNet and train the model on the dataset. 
Then we choose another three combinations of channels $[32, 32, 64, 128]$, $[32, 64, 128, 256]$, and $[64, 64, 128, 256]$ to train the LSNet. To design a lightweight model, the largest channel number is set to $256$. 
As shown in Fig. \ref{fig:loss}, the LSNet with $[16, 32, 64, 128]$ achieves the best performance in terms of train loss and test accuracy. 
As can be observed, except for the combination of $[64, 64, 128, 256]$, the other three combinations achieve similar performance in terms of test accuracy. 
However, to get a model with less parameters, we choose the combination of $[16, 32, 64, 128]$ as the final architecture of the LSNet.

\subsection{Performance under Federated Learning (FL) Condition}

To evaluate the performance of the proposed LSNet under the FL condition, we conduct experiments by varying the number of clients per round. 
As shown in Fig. \ref{fig:mrounds}, we set the number of clients as $5$, then during each round, we randomly select $[1, 2, 3, 4, 5]$ clients to participate in the training process. 
After $200$ rounds, the LSNet achieves the best performance when all $5$ clients participate in the training process, achieving an accuracy of $82.0\%$ for known UAVs. 
When only $1$ client participates in the training process, the LSNet achieves an accuracy of $72.55\%$, which is $9.45\%$ lower than the best performance. 
As for unknown UAVs, the LSNet achieves an AUROC of $0.68$ when all $5$ clients participate in the training process, which is $0.13$ higher than that when only $1$ client participates in the training process. 
This indicates that the LSNet is effective in detecting both known and unknown UAVs when more clients participate in the training process.

\begin{figure}[h!]
\centering
\subfigure[]{
\includegraphics[width=0.95\linewidth]{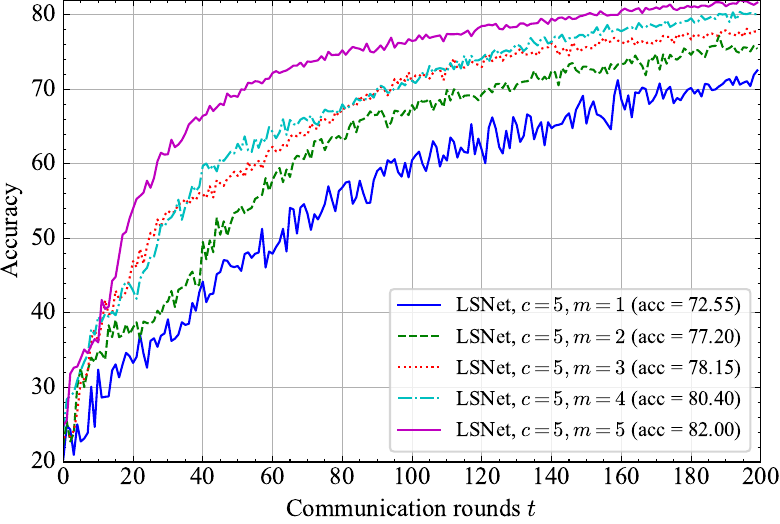}}
\subfigure[]{\includegraphics[width=0.95\linewidth]{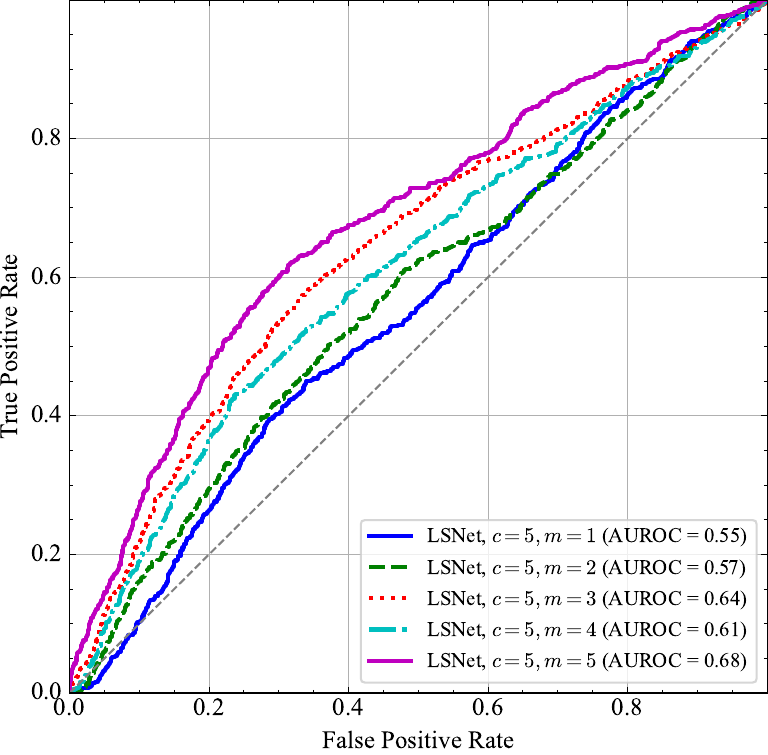}}
\DeclareGraphicsExtensions.
\caption{The performance of LSNet with different numbers of clients per round, (a) accuracy with round; (b) AUROC.}
\label{fig:mrounds}
\end{figure}

\begin{figure}
\centering
\subfigure[]{
\includegraphics[width=0.95\linewidth]{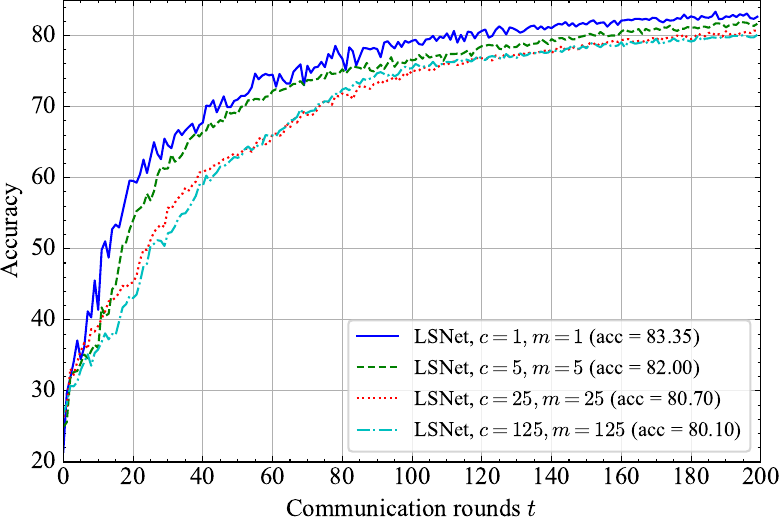}}
\subfigure[]{\includegraphics[width=0.95\linewidth]{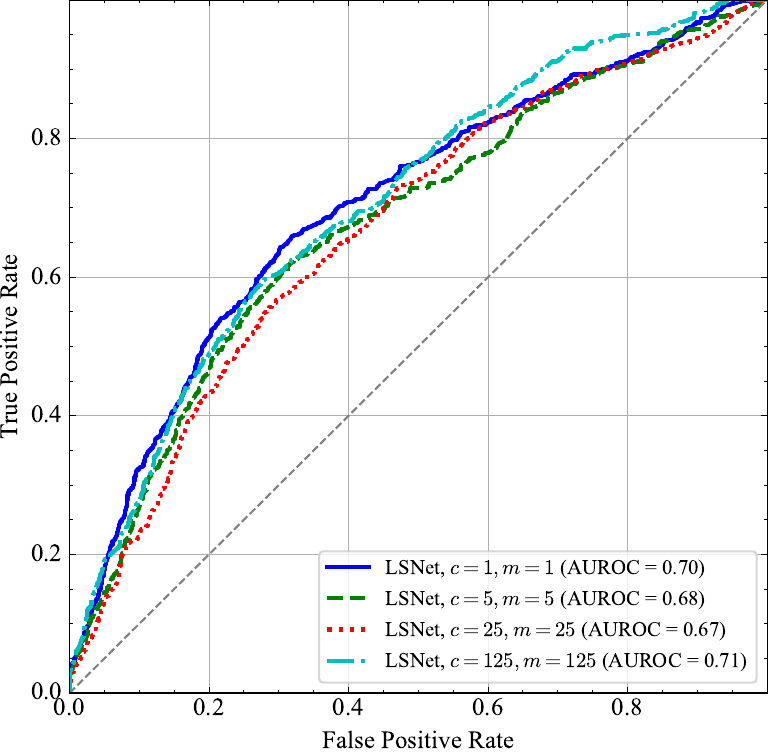}}
\DeclareGraphicsExtensions.
\caption{The performance of LSNet with different numbers of clinets, (a) accuracy with round; (b) AUROC.}
\label{fig:nclients}
\end{figure}

Moreover, we conduct experiments by varying the number of clients in the LSNet, as shown in Fig. \ref{fig:nclients}. 
We set the client numbers as $[1, 5, 25, 125]$ and train the LSNet on these different numbers of clients. 
As can be observed, the LSNet achieves the best performance when the number of clients is $1$, achieving an accuracy of $83.35\%$ for known UAVs. 
When the number of clients is $125$, the LSNet achieves an accuracy of $80.10\%$, which is $2.25\%$ lower than the best performance. 
As for unknown UAVs, the LSNet achieves an AUROC of $0.70$ when the number of clients is $1$, while it achieves $0.71$ when the number of clients is $125$. 
This indicates that the LSNet is effective in detecting both known and unknown UAVs when the number of clients is $1$. However, the performance of the LSNet is slightly affected when the number of clients is increased if all clients are used during training. 

\begin{figure}[h!]
\centering
\includegraphics[width=0.48\textwidth]{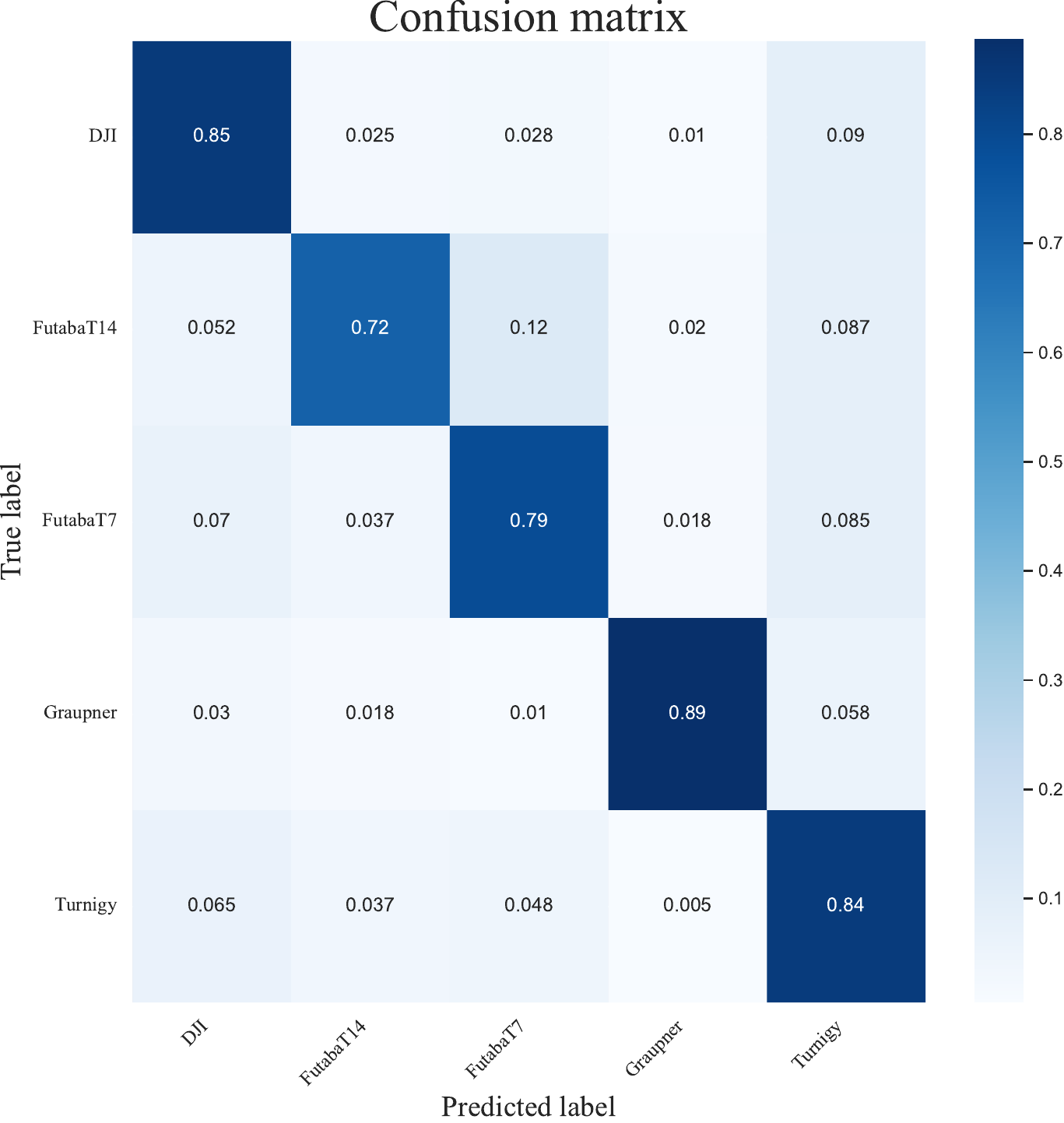}
\caption{Confusion matrix of known UAVs under $c=5, m=5$.}
\label{fig:confusion}
\end{figure}

To further demonstrate the effectiveness of the proposed LSNet under the ZT-FL condition, we present the confusion matrix of the LSNet under $c=5, m=5$ in Fig. \ref{fig:confusion}. 
As can be observed, for known UAVs, all UAVs are well classified except FutabaT14, with an accuracy of over $80\%$. 
Moreover, FutabaT14 and FutabaT7 are confused with each other, with an accuracy of $72\%$ and $79\%$, respectively. 
It is mainly because these two UAVs have similar RF signals (they are from the same manufacturer), which makes it difficult to distinguish them. 
This highlights a common challenge in UAV authentication systems where UAVs from the same manufacturer share similar transmission protocols, modulation schemes, and hardware components. To address this limitation, future research could explore enhanced feature extraction techniques that capture subtle differences in transmission patterns, or incorporate temporal analysis of signal sequences to identify manufacturer-specific but model-distinctive characteristics.

\subsection{Performance under Unknown UAV Situation}

To evaluate the performance of the LSNet under the unknown UAV situation, we conduct experiments by varying the number of unknown UAVs. 
The openness is calculated based on the number of known and unknown UAVs, as shown in Table \ref{tab:openness}. 
As shown in Fig. \ref{fig:mrounds}, we set the number of unknown UAVs as $[1, 2, 3, 4, 5, 6]$ and train the LSNet on these different numbers of unknown UAVs. 
Intuitively, the performance of the LSNet is affected by the number of unknown UAVs. 
The more unknown UAVs, the higher the performance of the LSNet in terms of accuracy and AUROC, and the lower the performance of the LSNet in terms of AUROC. However, when $2$ unknown UAVs are added, the performance of the LSNet is significantly affected, with an accuracy of $82.0\%$ for known UAVs and an AUROC of $0.68$ for unknown UAVs. This is mainly because the two unknown UAVs are Noise and Taranis, which are difficult to distinguish from known UAVs. Thus, for unknown UAVs, if the number of unknown UAVs is increased, the performance of the LSNet is significantly affected. Moreover, if unknown UAVs have similar RF signals to known UAVs, the performance of the LSNet is significantly affected. 

The ROC curves in Fig. \ref{fig:mrounds}(b) provide critical security performance metrics for practical UAV authentication applications. These curves illustrate the trade-off between the true positive rate (correctly identifying unknown UAVs) and false positive rate (incorrectly rejecting known UAVs). For security-critical applications, operating points toward the left of the ROC curves would be preferred, as they minimize false positives at the expense of potentially missing some unknown UAVs. For instance, with 1 unknown UAV type, at a false positive rate of 0.1, we achieve a true positive rate of approximately 0.55, meaning 55\% of unknown UAVs would be correctly rejected while only 10\% of known UAVs would be incorrectly rejected. As the number of unknown UAVs increases to 2, maintaining the same false positive rate of 0.1 reduces the true positive rate to approximately 0.35, highlighting the increasing difficulty in maintaining security as the diversity of potential threats grows. This security performance degradation is particularly evident when the system must distinguish between 6 unknown UAV types, where the AUROC approaches 0.50 (equivalent to random guessing), indicating that practically effective authentication would require accepting higher false positive rates or implementing additional security measures.

\begin{table}[h]
\centering
\caption{Openness for Different Numbers of Unknown Classes}
\begin{tabular}{ccc}
\toprule
\textbf{Number of Known} & \textbf{Number of Unknown} & \textbf{Openness} \\ \midrule
1 & 6 & 0.5000 \\ 
2 & 5 & 0.3333 \\ 
3 & 4 & 0.2254 \\ 
4 & 3 & 0.1472 \\ 
5 & 2 & 0.0871 \\ 
6 & 1 & 0.0392 \\ 
\bottomrule
\end{tabular}
\label{tab:openness}
\end{table}

\begin{figure}
\centering
\subfigure[]{
\includegraphics[width=0.95\linewidth]{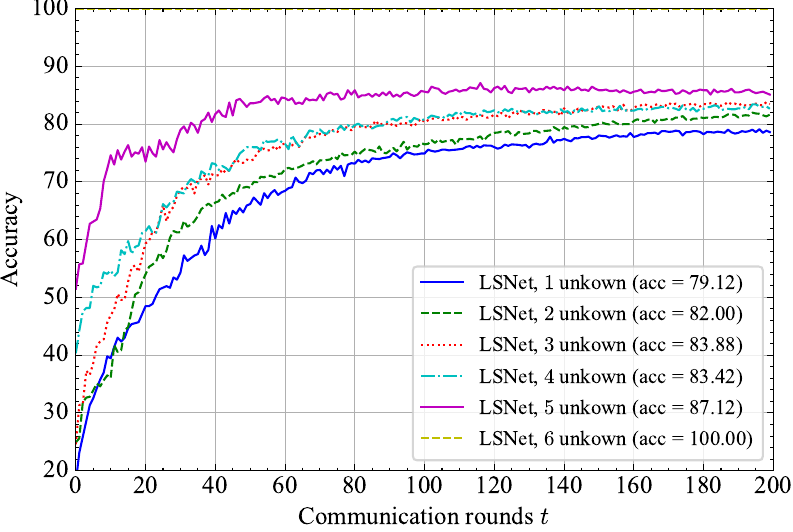}}
\subfigure[]{\includegraphics[width=0.95\linewidth]{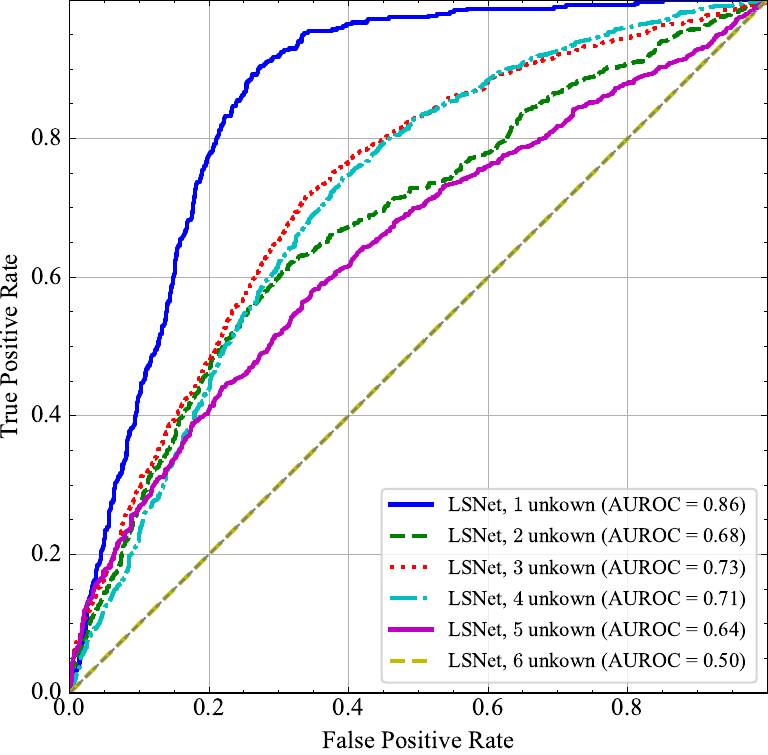}}
\DeclareGraphicsExtensions.
\caption{The performance of LSNet with different numbers of unknown UAVs, (a) accuracy with round; (b) AUROC.}
\label{fig:mrounds}
\end{figure}

\section{Conclusion}
\label{sec:conclusion}
In this paper, we present a novel federated learning (FL)-based lightweight spectrogram network (LSNet) with zero trust for UAV authentication and rejection. The proposed FL-based LSNet with zero trust can ensure the privacy and security of the data and wireless network during the model training process. The proposed LSNet is designed to extract features from RF signals and classify UAVs into known and unknown classes, by using multi-channel attention convolutional (MCAC) blocks and a novel class anchor loss. The experimental results demonstrate that the proposed LSNet outperforms the benchmark models in terms of accuracy, with fewer parameters and a smaller storage size. Moreover, the LSNet achieves the best performance when all $5$ clients participate in the training process, achieving an accuracy of over $80\%$ for known UAVs and an AUROC of $0.7$ for unknown UAVs. Additionally, the effect of the number of clients per round, the number of clients, and the number of unknown UAVs are investigated. These results indicate that our framework has considerable practical significance in FL situations.

\bibliographystyle{IEEEtran}
\bibliography{fed_uav}

\begin{IEEEbiography}[{\includegraphics[width=1in,height=1.25in,clip,keepaspectratio]{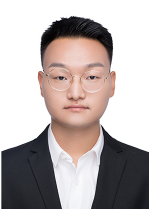}}]{Hao Zhang} (Member, IEEE) received the B.E. and M.Eng. from Nanchang University, China, in 2017 and 2020, respectively, and the Ph.D. degree from Nanjing University of Aeronautics and Astronautics, Nanjing, China, in 2025. He is now a postdoctoral fellow with the College of Artificial Intelligence, Nanjing University of Aeronautics and Astronautics, China. He was a visiting Ph.D. student at the School of Electrical \& Electronic Engineering, Nanyang Technological University, Singapore, in 2024. His research interests focus on deep learning, foundation models, wireless communication, signal processing, and spectrum cognition.
\end{IEEEbiography}

\vspace{-4em}

\begin{IEEEbiography}[{\includegraphics[width=1in,height=1.25in,clip,keepaspectratio]{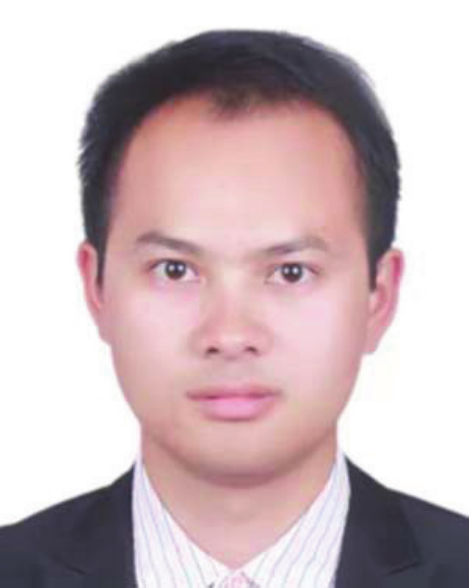}}]{Fuhui Zhou} (Senior member, IEEE) s currently a Full Professor with Nanjing University of Aeronautics and Astronautics, Nanjing, China, where he is also with the Key Laboratory of Dynamic Cognitive System  of  Electromagnetic  Spectrum  Space.  His research interests include cognitive radio, cognitive intelligence, knowledge graph, edge computing, and resource allocation.

Prof.  Zhou  has  published   over  200  papers   in internationally renowned journals and conferences in the field of communications. He has been selected for  1  ESI  hot  paper  and   13  ESI  highly  cited  papers.  He  has  received  4 Best Paper Awards at international conferences such as IEEE Globecom and IEEE  ICC.  He  was  awarded  as  2021  Most  Cited  Chinese  Researchers  by Elsevier,  Stanford  World’s  Top  2\%  Scientists,  IEEE  ComSoc  Asia-Pacific Outstanding Young Researcher and Young Elite Scientist Award of China and URSI GASS Young Scientist. He serves as an Editor of IEEE Transactions on  communication,  IEEE  Systems  Journal,  IEEE  Wireless  Communication Letters, IEEE Access and Physical Communications.
\end{IEEEbiography}
\vspace{-4em}

\begin{IEEEbiography}[{\includegraphics[width=1in,height=1.25in, clip,keepaspectratio]{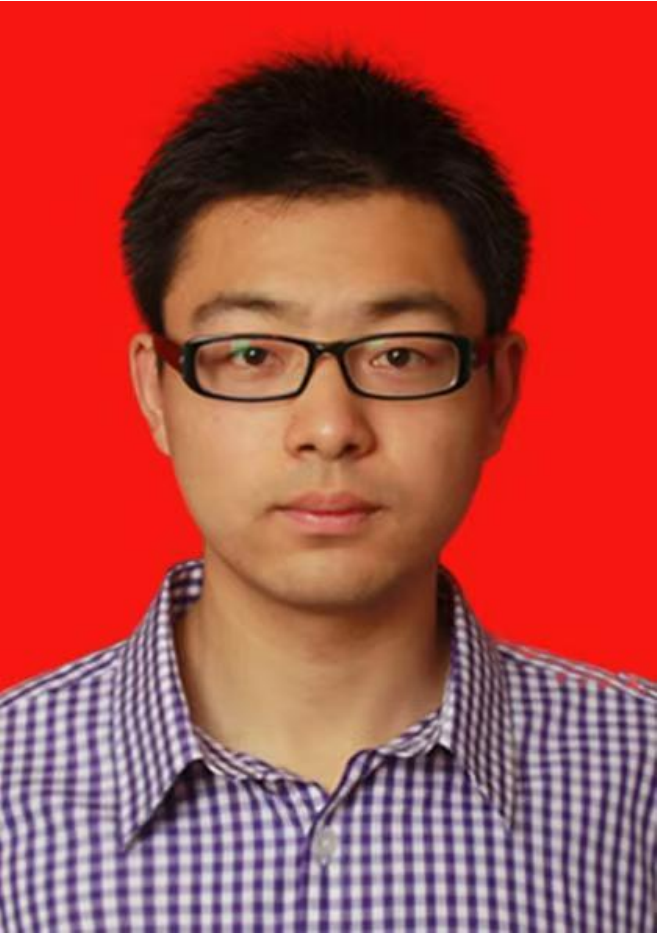}}]{Wei Wang} (S'10-M'16-SM'20) is currently a full professor with the School of Computer Science, Wuhan University. He received the Ph.D. degree from the Department of Computer Science and Engineering, The Hong Kong University of Science and Technology. His research interests include PHY/MAC design and mobile computing in wireless systems. He has published 2 books and over 100 refereed papers in international leading journals and primer conferences. He is the inventor of 3 US and 20 Chinese patents. He won the best paper award in IEEE ICC 2019. He was selected as Young Elite Scientist Sponsorship Program, China Association for Science and Technology, and Hundred Talents Program, Hubei Province, China. He served on TPC of INFOCOM, GBLOBECOM, ICC, etc. He served as Editors for IEEE TMC, ACM IMWUT.
\end{IEEEbiography}
		
\vspace{-4em}

\begin{IEEEbiography}[{\includegraphics[width=1in,height=1.25in,clip,keepaspectratio]{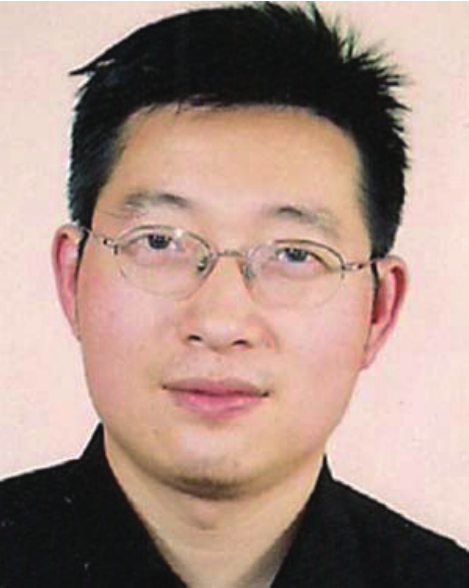}}]{Qihui Wu}
(Fellow, IEEE) received the B.S. degree in communications engineering, the M.S. and Ph.D. degrees in communications and information systems from the Institute of Communications Engineering, Nanjing, China, in 1994, 1997, and 2000, respectively. From 2003 to 2005, he was a Postdoctoral Research Associate with Southeast University, Nanjing, China. From 2005 to 2007, he was an Associate Professor with the College of Communications Engineering, PLA University of Science and Technology, Nanjing, China, where he was a Full Professor from 2008 to 2016. Since May 2016, he has been a Full Professor with the College of Electronic and Information Engineering, Nanjing University of Aeronautics and Astronautics, Nanjing, China. From March 2011 to September 2011, he was an Advanced Visiting Scholar with the Stevens Institute of Technology, Hoboken, USA. His current research interests span the areas of wireless communications and statistical signal processing, with emphasis on system design of software defined radio, cognitive radio, and smart radio.
\end{IEEEbiography}

\begin{IEEEbiography}[{\includegraphics[width=1in,height=1.25in,clip,keepaspectratio]{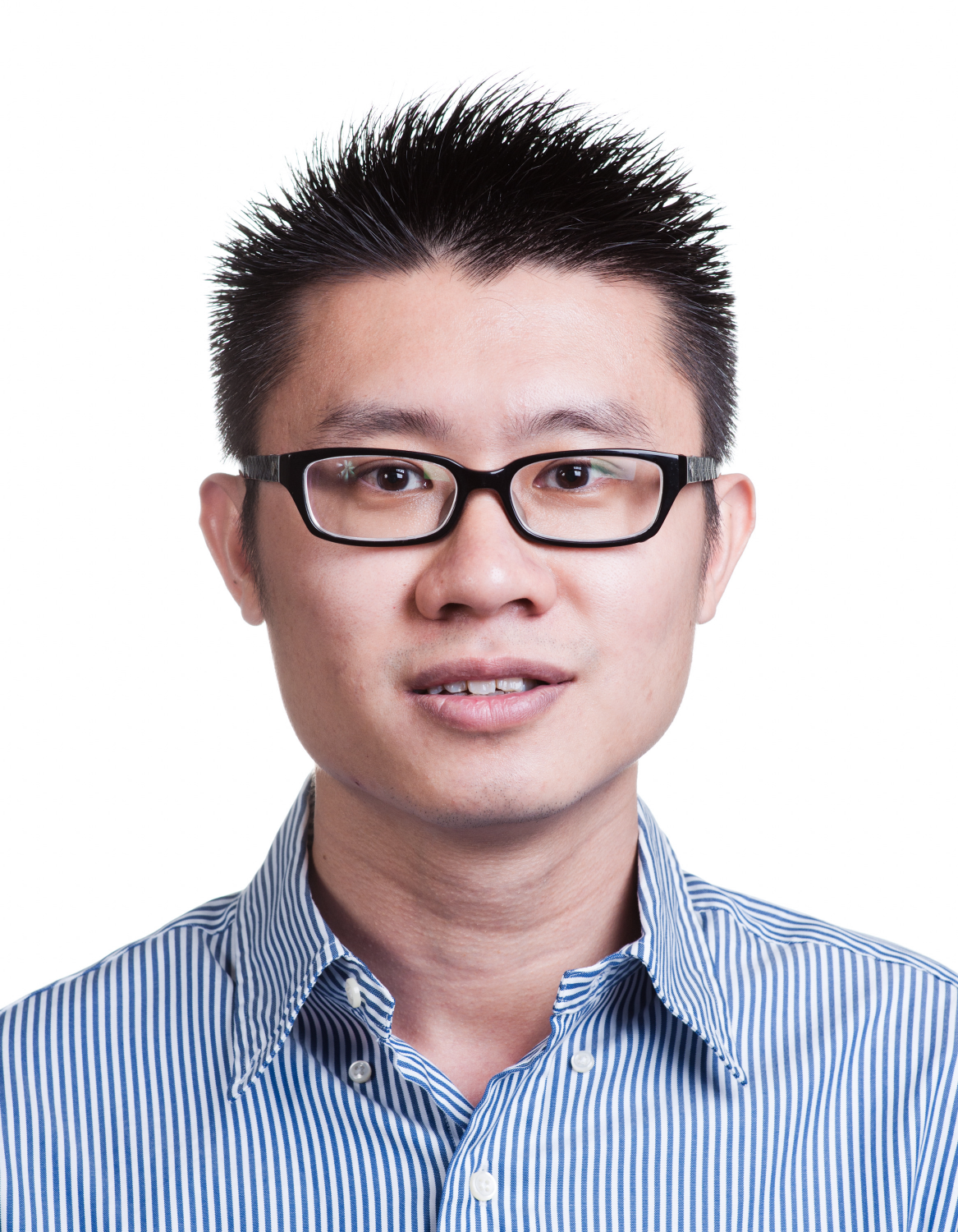}}]{Chau Yuen}
(S'02-M'06-SM'12-F'21) received the B.Eng. and Ph.D. degrees from Nanyang Technological University, Singapore, in 2000 and 2004, respectively. He was a Post-Doctoral Fellow with Lucent Technologies Bell Labs, Murray Hill, in 2005. From 2006 to 2010, he was with the Institute for Infocomm Research, Singapore. From 2010 to 2023, he was with the Engineering Product Development Pillar, Singapore University of Technology and Design. Since 2023, he has been with the School of Electrical and Electronic Engineering, Nanyang Technological University, currently he is Provost’s Chair in Wireless Communications, Assistant Dean in Graduate College, and Cluster Director for Sustainable Built Environment at ER@IN.
 
Dr. Yuen received IEEE Communications Society Leonard G. Abraham Prize (2024), IEEE Communications Society Best Tutorial Paper Award (2024), IEEE Communications Society Fred W. Ellersick Prize (2023), IEEE Marconi Prize Paper Award in Wireless Communications (2021), IEEE APB Outstanding Paper Award (2023), and EURASIP Best Paper Award for JOURNAL ON WIRELESS COMMUNICATIONS AND NETWORKING (2021).
 
Dr Yuen current serves as an Editor-in-Chief for Springer Nature Computer Science, Editor for IEEE TRANSACTIONS ON VEHICULAR TECHNOLOGY, IEEE TRANSACTIONS ON NEURAL NETWORKS AND LEARNING SYSTEMS, and IEEE TRANSACTIONS ON NETWORK SCIENCE AND ENGINEERING, where he was awarded as IEEE TNSE Excellent Editor Award 2024 and 2022, and Top Associate Editor for TVT from 2009 to 2015. He also served as the guest editor for several special issues, including IEEE JOURNAL ON SELECTED AREAS IN COMMUNICATIONS, IEEE WIRELESS COMMUNICATIONS MAGAZINE, IEEE COMMUNICATIONS MAGAZINE, IEEE VEHICULAR TECHNOLOGY MAGAZINE, IEEE TRANSACTIONS ON COGNITIVE COMMUNICATIONS AND NETWORKING, and ELSEVIER APPLIED ENERGY.
 
He is listed as Top 2\% Scientists by Stanford University, and also a Highly Cited Researcher by Clarivate Web of Science from 2022. He has 4 US patents and published over 500 research papers at international journals.
\end{IEEEbiography}

\end{document}